\documentclass[aps,pre,reprint,showpacs]{revtex4-1}

\usepackage{verbatim}
\usepackage{graphicx}
\usepackage{amsmath}
\usepackage{amssymb}
\usepackage[usenames, dvipsnames]{color}
\usepackage[normalem]{ulem}
\begin{document}

\preprint{}

\title{Optimal Control of Overdamped Systems}

\author{Patrick R. Zulkowski}
\email[]{pzulkowski@berkeley.edu}
\affiliation{Department of Physics, University of California, Berkeley, CA 94720}
\affiliation{Department of Mathematics, Berkeley City College, Berkeley, California 94704, USA}
\affiliation{Redwood Center for Theoretical Neuroscience, University of California, Berkeley CA 94720}

\author{Michael R. DeWeese}
\email[]{deweese@berkeley.edu}
\affiliation{Department of Physics, University of California, Berkeley, CA 94720}
\affiliation{Redwood Center for Theoretical Neuroscience, University of California, Berkeley CA 94720}
\affiliation{Helen Wills Neuroscience Institute, University of California, Berkeley CA 94720}

\begin{abstract}
Nonequilibrium physics encompasses a broad range of natural and synthetic small-scale systems. Optimizing transitions of such systems will be crucial for the development of nanoscale technologies and may reveal the physical principles underlying biological processes at the molecular level. Recent work has demonstrated that when a thermodynamic system is driven away from equilibrium then the space of controllable parameters has a Riemannian geometry induced by a generalized inverse diffusion tensor. We derive a simple, compact expression for the inverse diffusion tensor that depends solely on equilibrium information for a broad class of potentials. We use this formula to compute the minimal dissipation for two model systems relevant to small-scale information processing and biological molecular motors. In the first model, we optimally erase a single classical bit of information modelled by an overdamped particle in a smooth double-well potential. In the second model, we find the minimal dissipation of a simple molecular motor model coupled to an optical trap. In both models, we find that the minimal dissipation for the optimal protocol of duration $ \tau $ is proportional to $ 1 / \tau $, as expected, though the dissipation for the erasure model takes a different form than what we found previously for a similar system.  \end{abstract}


\date{\today}

\maketitle

\section{Introduction}
A complete set of principles describing physical phenomena far from equilibrium remains elusive. However, considerable progress has been made in the study of nonequilibrium processes in recent times. Fluctuation theorems relating the probability of an increase to that of a comparable decrease in entropy during a finite time period have been derived~\cite{Evans1993,Evans1994,Gallavotti1995a,Crooks1999,Hatano2001} and experimentally verified~\cite{Wang2002,Carberry2004,Garnier2005,Toyabe2010}. Considerations of Maxwell's demon and Landauer's principle have led to a better understanding of the thermodynamic role of information~\cite{Szilard1929,Landauer1961,Bennett1982} and new fundamental relationships valid for systems far from equilibrium such as the Jarzynski equality have provided deep insights into thermodynamic quantities such as
%like 
entropy~\cite{Jarzynski1997,Liphardt2002,Seifert2005b,Sagawa2010}. More recently it has been appreciated that these nonequilibrium relationships are closely related to pioneering work by Bochkov and Kuzovlev~\cite{Bochkov_1977,Bochkov_2_1977,Bochkov_1981,Bochkov_2_1981,Bochkov2013,Jarzynski_comparison_2007,Seifert_Stochastic_2008,Campisi_2011,Horowitz_2007}. 

Recent work has also shed light on the general problem of computing optimal protocols that minimize dissipation while driving small-scale systems between stationary states~\cite{Sivak2012,Zulkowski2012,Zulkowski_PLOS_ONE_2013,Shenfeld2009,Brody2009,Seifert2008,Seifert2007,Aurell2011}. Optimization schemes for finite-time thermodynamic processes will be needed for technological applications in which energetic efficiency is paramount~\cite{Andresen2011,Chen2004}. This will be particularly relevant in the decades to come as computational demands approach physical limits.

It may also be the case that evolution sculpted molecular machines such as kinesin and $\text{F}_{\text{O}}-\text{F}_{1} $ ATP-ase to operate far from equilibrium while maximizing efficiency~\cite{MBOC}. Thus, optimization schemes for nonequilibrium transitions also have the potential to unlock physical principles underlying the function of biological systems on the molecular level.

A promising approach utilizes a linear-response framework in which a generalized inverse diffusion tensor induces a Riemannian manifold structure on the space of parameters~\cite{Sivak2012}. Optimal protocols are geodesics of this geometry. This idea is developed further in~\cite{Zulkowski2012} where the machinery of Riemannian geometry is exploited to find explicit optimal protocols for a paradigmatic colloidal particle model and in~\cite{Zulkowski2014} where optimal finite-time erasure protocols are computed for a simple classical bit model. An extension of this framework to transitions between nonequilibrium steady states is established in~\cite{Zulkowski_PLOS_ONE_2013}. 

This framework builds on earlier investigations into thermodynamic metrics by including the dynamics of the driven system. For macroscopic systems, the properties of optimal driving processes have been investigated using thermodynamic length, a natural measure of the distance between pairs of equilibrium thermodynamics states \cite{Weinhold1975a,Ruppeiner1979,Schlogl1985,Salamon1984,Salamon1983a,Brody1995}, with extensions to microscopic systems involving a metric of Fisher information \cite{Crooks2007c,Burbea1982}. Slow transitions between nonequilibrium steady states have also been studied in terms of thermodynamic metric structure~\cite{Mandal2015}.

In~\cite{Zulkowski2012}, the method to calculate the inverse diffusion tensor components relied on the (continuous) potential being harmonic. In this paper, we demonstrate how the inverse diffusion tensor may be computed for a more general class of potentials in terms of the equilibrium probability distribution. Our starting point will be the Fokker-Planck equation~\cite{Zwanzig2001}, which we assume fully describes the physics of the system. Furthermore, we show that the inverse diffusion tensor arises naturally through an expansion in temporal derivatives in this general setting. 

We use this construction to compute minimal  dissipation for two model systems of physical interest. First, we consider a one-dimensional system modeling the storage and erasure of a single classical bit of information~\cite{Lutz2009,Lutz2012,Zulkowski2014}. The erasure of information results in energy dissipation according to the Landauer principle. Minimizing this dissipation (equivalently, maximizing erasure efficiency) will likely prove critical to the development of future small-scale information processing devices. The first model system consists of an overdamped colloidal particle diffusing under the influence of a continuous double-well potential with a large central barrier stabilizing the memory. If the particle is found to the left (right) of the origin, the memory value is 1 (0). We seek the most efficient protocol altering the shape of the confining potential so that the particle 
will be found to the right of the origin with overwhelming probability, thus setting the memory value to 0 and erasing the single bit of classical information originally encoded by the system.

\begin{figure}
\begin{center}
\includegraphics{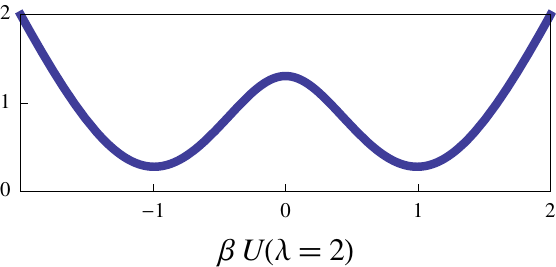} \\
\includegraphics{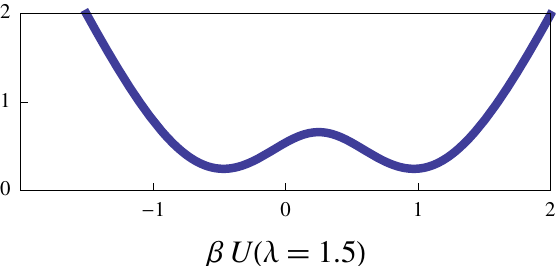} \\
\includegraphics{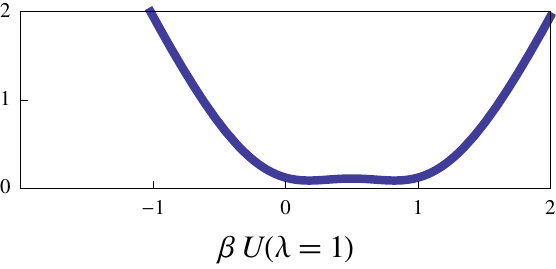} \\
\includegraphics{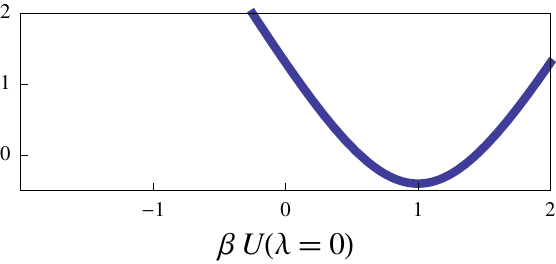} \\
\end{center}
\caption{Continuous erasure protocol. The lefthand well of the double-well potential merges with the right and the central barrier lowers simultaneously as $ \lambda $ decreases from $ 2 $ to $ 0 $.}
\label{fig:erasure}
\end{figure}

The erasure cycle consists of a continuous stage in which the wells merge and the central barrier is lowered. A reset stage in which the potential returns instantaneously to its original state and leaves the final probability distribution undisturbed completes the erasure cycle (see Fig.~\ref{fig:erasure}). The inverse diffusion tensor predicts optimal erasure cycles in the long duration limit. 

The second model system consists of an overdamped colloidal particle diffusing in one-dimension while coupled to a ratchet potential and an optical trap. In this simplified model, the coordinate of the diffusing particle may be identified with a mechanical state variable of a molecular motor~\cite{Reimann2002} and the thermal bath consists of the huge number of irrelevant degrees of freedom of the liquid surrounding the motor as well as the internal degrees of freedom of the motor itself and the structures with which it interacts. We suppose that our simplistic molecular motor couples to an optical trap. The inverse diffusion tensor framework predicts the optimal time course for the optical trap center, which represents an external drive for this simple molecular machine.
%driving of this simplistic molecular machine. 

\section{The Inverse Diffusion Tensor}

For a physical system at equilibrium in contact with a thermal bath, the probability distribution over microstates $ x $ is given by the canonical ensemble
\begin{equation}\label{canonical} \rho_{eq}( x ,\boldsymbol \lambda) \equiv \exp{ \beta \left[ F(\boldsymbol \lambda) - E(x,\boldsymbol \lambda) \right] } \ ,  \end{equation}
where $ \beta = (k_{\rm B} T)^{-1} $ is the inverse temperature in natural units, $ F(\boldsymbol \lambda) $ is the free energy, and $ E(x,\boldsymbol \lambda) $ is the system energy as a function of the microstate $ x $ and a collection of experimentally controllable parameters $ \boldsymbol \lambda $.

In equilibrium, the thermodynamic state of the system (the probability distribution over microstates) is completely specified by values of the control parameters, but out of equilibrium the system's probability distribution over microstates fundamentally depends on the history of the control parameters $\boldsymbol\lambda $, which we denote by the control parameter protocol $\boldsymbol \Lambda$. We assume the protocol to be sufficiently smooth to be twice-differentiable.

The average excess power exerted by the external agent on the system, over and above the average power on a system at equilibrium, is~\cite{Zulkowski2012}
\begin{equation}\label{expower} \beta(t_{0}) \mathcal{P}_{\rm ex}(t_{0}) \equiv - \left[ \frac{ d \boldsymbol \lambda^{T} }{d t} \right]_{t_{0}} \cdot \left< \delta \boldsymbol X \right>_{ \boldsymbol \Lambda} \ . \end{equation}
Here $ \boldsymbol X \equiv -\frac{ \partial \left( \beta E \right)}{\partial \boldsymbol \lambda} $ are the forces conjugate to the control parameters $ \boldsymbol \lambda $, and $ \delta \boldsymbol X (t_{0}) \equiv \boldsymbol X(t_{0}) - \left< \boldsymbol X \right>_{\boldsymbol \lambda(t_{0}) } $ is the deviation of ${\bf X}(t_0)$ from its current equilibrium value.

Applying linear response theory~\cite{Zwanzig2001},
\begin{equation} \left< \delta \boldsymbol  X(t_{0}) \right>_{ \boldsymbol \Lambda} \approx \int_{-\infty}^{t_{0}} \boldsymbol \chi(t_{0}-t') \cdot \left[ \boldsymbol \lambda(t_{0}) - \boldsymbol \lambda(t') \right] dt' \end{equation}
where $ \chi_{ij}(t) \equiv - d\Sigma_{ij}^{\boldsymbol \lambda(t_{0})}(t)/dt $ represents the response
of conjugate force $ X_{i} $ at time $ t $ to a perturbation in control
parameter $ \lambda^{j} $ at time zero, and
\begin{equation} \Sigma_{ij}^{\boldsymbol \lambda(t_{0})}(t) \equiv \left< \delta X_{j}(0) \delta X_{i}(t) \right>_{\boldsymbol \lambda(t_{0})}. \end{equation}
For protocols that vary sufficiently slowly~\cite{Sivak2012b}, the resulting mean excess power is
\begin{equation}\label{etadef} \beta(t_{0}) \mathcal{P}_{\rm ex}(t_{0}) \approx  \left[ \frac{ d \boldsymbol \lambda^{T} }{d t}  \right]_{t_{0}} \cdot \boldsymbol \zeta (\boldsymbol \lambda(t_{0}) ) \cdot \left[ \frac{ d \boldsymbol \lambda }{d t}  \right]_{t_{0}} \ , \end{equation}
for inverse diffusion tensor
\begin{equation} \zeta_{ij} \equiv  \int_{0}^{\infty} dt' \big\langle \delta X_{j}(0) \delta X_{i}(t') \big\rangle_{\boldsymbol \lambda(t_{0})}\ ~, \end{equation}
or, more conveniently, 
\begin{equation}\label{eq:zetacomps} \zeta_{ij} =  \int_{0}^{\infty} dt' \big\langle \partial_{\lambda^j}\phi(0) ~\partial_{\lambda^i}\phi(t') \big\rangle_{\boldsymbol \lambda(t_{0})}~,\ \end{equation}
where $ \phi(x,\boldsymbol \lambda) \equiv - \ln \rho_{eq}(x,\boldsymbol \lambda) $.

We assume our system consists of an overdamped colloidal particle obeying the stochastic equation of motion
\begin{equation}\label{eq:brownian} \dot{x} = -\frac{1}{\gamma} \partial_{x}U(x(t),t)+ F(t) \end{equation}
for Gaussian white noise $ F(t) $ satisfying
\begin{equation} \langle F(t) \rangle = 0 \ , \ \langle F(t) F(t') \rangle = \frac{2}{\beta \gamma} \delta(t-t'). \end{equation}
Here, $ \gamma $ is the Cartesian friction coefficient and $ U(x,t) $ is a generic potential growing unbounded as $ |x| \rightarrow \infty $ at a rate to be specified shortly.

The components in Eq.~\eqref{eq:zetacomps} may be readily computed if Eq.~\eqref{eq:brownian} is linear or, equivalently, the potential is harmonic~\cite{Zulkowski2012}. The linearity of Eq.~\eqref{eq:brownian} allows us to write general solutions as a linear combination of a homogeneous piece dependent only on the initial conditions and a particular piece dependent on the Gaussian noise. Such a decomposition allows for a straightforward calculation of the time correlation functions appearing in Eq.~\eqref{eq:zetacomps} as demonstrated in~\cite{Zulkowski2012}.

For more general potentials, such a decomposition is impossible and so a different approach must be found to compute Eq.~\eqref{eq:zetacomps}. We will find the equivalent statistical description in terms of the Fokker-Planck equation
\begin{equation}\label{eq:FP} \partial_{t} \rho = D \big[ \partial_{x} \left( \beta U'(x,t) \rho \right) + \partial_{x}^2 \rho \big] \equiv - \partial_{x} G \end{equation}
convenient, where $ \rho(x,t) $ is the position probability density, $ G(x,t) $ is the probability current, and $ D $ is the diffusion coefficient.

We set out to compute a general expression for the inverse diffusion tensor of a driven system obeying overdamped dynamics. Using the method of Laplace transform~\cite{Pankratov1999}, we succeeded in writing the components of the tensor entirely in terms of the equilibrium probability distribution $ \rho_{eq} $ and cumulative distribution function $ \Pi_{eq} $:
\begin{equation}\label{eq:zetaPi} \zeta_{ij}(\boldsymbol \lambda) = \frac{1}{D} \int_{-\infty}^{\infty} dx ~  \bigg[ \frac{\partial_{\lambda^{i}} \Pi_{eq}(x, \boldsymbol \lambda) ~\partial_{\lambda^{j}} \Pi_{eq}(x , \boldsymbol \lambda)}{\rho_{eq}(x,\boldsymbol \lambda)} \bigg]. \end{equation}

We assume that the potential satisfies $ U(x, \boldsymbol \lambda) \rightarrow \infty $ as $ |x| \rightarrow \infty $. Note that the construction also applies for reflecting-wall boundary conditions.

The inverse diffusion tensor components are given by
\begin{equation}\label{eq:misc-tensor} \zeta_{ij}(\boldsymbol \lambda) = \int_{0}^{\infty} dt'~\bigg< \frac{\partial \phi}{\partial \lambda^{i}}(t') \frac{\partial \phi}{\partial \lambda^{j}}(0)  \bigg>_{eq,\boldsymbol \lambda} , \end{equation}
where $ \phi(x,\boldsymbol \lambda) \equiv - \ln \rho_{eq}(x, \boldsymbol \lambda) $ and $ \rho_{eq}(x,\boldsymbol \lambda) $ is the equilibrium distribution.

We rewrite Eq.~\eqref{eq:misc-tensor} as
\begin{align}\label{eq:misc-expandcorr}
\zeta_{ij}(\boldsymbol \lambda) = \int_{0}^{\infty} dt'~& \bigg[  \int_{-\infty}^{\infty} dx_{0}~ \rho_{eq}(x_{0} , \boldsymbol \lambda)~\partial_{\lambda^j} \phi(x_{0} , \boldsymbol \lambda) \times  \nonumber \\ & \bigg( \int_{-\infty}^{\infty} dx~\rho(x,t'; x_{0})~\partial_{\lambda^i} \phi(x, \boldsymbol \lambda) \bigg) \bigg],
\end{align}
where $ \rho(x,t; x_{0}) $ satisfies Eq.~\eqref{eq:FP} with initial condition $ \rho(x,t=0; x_{0}) = \delta_{x,x_{0}} $ and $ \rho(x,t;x_{0}) \rightarrow 0 $ for $ |x| \rightarrow \infty $. For simplicity, define 
\begin{equation} m_{i}(t) \equiv \int_{-\infty}^{\infty} dx~\rho(x,t; x_{0})~\partial_{\lambda^i} \phi(x,\boldsymbol \lambda) \end{equation}
 so that
\begin{equation}\label{eq:misc-simpcomps} \zeta_{ij}(\boldsymbol \lambda) =  \int_{-\infty}^{\infty}  dx_{0}~ \rho_{eq}(x_{0} , \boldsymbol \lambda)~\partial_{\lambda^j} \phi(x_{0} , \boldsymbol \lambda)~\int_{0}^{\infty} dt'~m_{i}(t'). \end{equation}
Note that we have suppressed the dependence of $ m_{i} $ on $ x_{0} $ and $ \boldsymbol \lambda $ for convenience.
We evaluate $ \int_{0}^{\infty} dt'~m_{i}(t') $ by computing the Laplace transform 
\begin{equation} \hat{m}_{i}(s) \equiv \int_{0}^{\infty} dt'~m_{i}(t')~e^{-s t'} \end{equation}
and taking the limit as $ s \rightarrow 0^{+} $. 

Integrating by parts,
\begin{equation} \int_{0}^{\infty} dt'~\frac{dm_{i}}{dt'}(t')~e^{-s t'} = s~\hat{m}_{i}(s)-m_{i}(0). \end{equation}
Note that $ m_{i}(\infty) $ vanishes since $ \lim_{t \rightarrow \infty} \rho(x,t; x_{0}) = \rho_{eq}(x,\boldsymbol \lambda) $; \emph{i.e.} the system comes to a stationary state after a sufficiently long time has elapsed. By definition of $ m_{i} $,
\begin{equation} m'_{i}(t) = \int_{-\infty}^{\infty} dx~\partial_{t}\rho(x,t; x_0)~\partial_{\lambda^i} \phi(x , \boldsymbol \lambda). \end{equation}
In terms of the probability current $ G(x,t) $,
\begin{equation}\label{eq:misc-Laplacem} \hat{m}_{i}(s) = \frac{m_{i}(0) -\int_{-\infty}^{\infty} dx~\partial_{x}\hat{G}(x,s)~\partial_{\lambda^i} \phi(x , \boldsymbol \lambda)}{s}. \end{equation}
Therefore, to compute $ \hat{m}_{i}(s) $, we need the Laplace transform of the probability current.

The Fokker-Planck equation may be used to derive an equation for the probability current:
\begin{equation} \partial_{t}G(x,t) = D \big[\beta U'(x,\boldsymbol \lambda) \partial_{x} G(x,t) + \partial_{x}^2 G(x,t) \big]. \end{equation}
Taking the Laplace transform of both sides, we have
\begin{equation} s~\hat{G}(x,s)-G(x,0) = D \big[\beta U'(x,\boldsymbol \lambda)~\partial_{x} \hat{G}(x,s) + \partial_{x}^2 \hat{G}(x,s) \big] \end{equation}
which follows from $ \lim_{t\rightarrow \infty} G(x,t) = 0 $. Multiplying both sides by $ s $ and defining $ H(x,s) \equiv s~\hat{G}(x,s) $,
\begin{equation}\label{eq:misc-H} s~H(x,s) - s~G(x,0) = D \big[\beta U'(x,\boldsymbol \lambda)~\partial_{x} H(x,s) + \partial_{x}^2 H(x,s) \big]. \end{equation}
We may obtain a solution to Eq.~\eqref{eq:misc-H} by expanding $ H(x,s) $ as a series in $ s$. If we define $H(x,s) \equiv H_{0}(x)+ s~H_{1}(x)+ \dots $, then
\begin{equation}\label{eq:misc-Hseq0} 0 = \beta U'(x,\boldsymbol \lambda)~H_{0}'(x) + H_{0}''(x) \end{equation}
\begin{equation} \label{eq:misc-Hseq1} H_{0}(x)-G(x,0) = D \big[\beta U'(x,\boldsymbol \lambda)~\partial_{x} H_1(x) + \partial_{x}^2 H_1(x) \big] \end{equation}
\begin{equation}\label{eq:misc-Hseq2} H_{k-1}(x) = D \big[\beta U'(x,\boldsymbol \lambda)~\partial_{x} H_{k}(x) + \partial_{x}^2 H_{k}(x) \big] \end{equation}
follow from substituting the expansion into Eq.~\eqref{eq:misc-H} and comparing the coefficients of powers of $ s $ on both sides. The boundary conditions on the probability current must also be satisfied by $ H_{k} $ for each $ k $. 

We see that these differential equations may be solved iteratively. Fortunately, it turns out that only $ H_{2}(x) $ is needed for our purposes as a short calculation using Eq.~\eqref{eq:misc-Laplacem} shows that
\begin{equation}\label{eq:intm} \int_{0}^{\infty} dt'~m_{i}(t') = - \int_{-\infty}^{\infty}  dx~\partial_{x}H_{2}(x)~\partial_{\lambda^i} \phi(x,\boldsymbol \lambda). \end{equation}

For potentials that grow unbounded as $ |x| \rightarrow \infty $, the probability current must vanish in the limit of large $ |x| $. With these boundary conditions it is not difficult to show
\begin{align}\label{eq:H2} H_{2}(x)  = \frac{1}{D} & \bigg[ - \bigg( \int_{-\infty}^{\infty} dx \ e^{- \beta U(x, \boldsymbol \lambda)} \int_{a}^{x} dx' \ e^{\beta U(x', \boldsymbol \lambda)} \times \nonumber \\ & \big( \theta(x'-x_{0}) - \Pi_{eq}(x', \boldsymbol \lambda) \big) \bigg) \Pi_{eq}(x, \boldsymbol \lambda) + \nonumber \\ &\int_{-\infty}^{x} dx' \ e^{- \beta U(x', \boldsymbol \lambda)} \int_{a}^{x'} dx'' \ e^{\beta U(x'', \boldsymbol \lambda)} \times \nonumber \\ &  \big( \theta(x''-x_{0}) - \Pi_{eq}(x'', \boldsymbol \lambda) \big) \bigg], \end{align}
where $ \theta $ denotes the Heaviside function and $ \Pi_{eq}(x, \boldsymbol \lambda) = \int_{-\infty}^{x} dx'~\rho_{eq}(x', \boldsymbol \lambda) $ is the equilibrium cumulative distribution function. Here, $ a $ is an arbitrary real constant. Surprisingly, $ H_{2}(x) $ is independent of $ a $ and we will have occasion to choose different convenient values for computational purposes.

From this result we see that the inverse diffusion tensor has the compact form
\begin{align}\label{eq:misc-compacttensor} \zeta_{ij}(\boldsymbol \lambda)  = \int d \nu(x,x',x'') \ & \bigg[ e^{\beta \left( U(x',\boldsymbol \lambda)-U(x,\boldsymbol \lambda)-U(x'',\boldsymbol \lambda) \right)}  \times \nonumber \\ & \partial_{\lambda^i}\phi(x,\boldsymbol \lambda) ~\partial_{\lambda^j}\phi(x'', \boldsymbol \lambda) \bigg], \end{align}
where we have used the shorthand
\begin{equation} \int d\nu(x,x',x'') \rightarrow -\frac{1}{D Z(\boldsymbol \lambda)} \int_{-\infty}^{\infty} dx \ \int_{a}^{x} dx' \ \int_{-\infty}^{x'} dx''. \end{equation}

This expression may be further simplified by observing that
\begin{equation} \int_{-\infty}^{x'} dx''~\partial_{\lambda^{j}} \phi(x'', \boldsymbol \lambda) e^{-\beta U(x'',\boldsymbol \lambda)} = -Z(\boldsymbol \lambda) \partial_{\lambda^{j}} \Pi_{eq}(x',\boldsymbol \lambda) \end{equation}
and
\begin{equation} \partial_{\lambda^{i}} \phi(x, \boldsymbol \lambda) e^{-\beta U(x, \boldsymbol \lambda)} = -Z(\boldsymbol \lambda) \partial_{\lambda^{i}} \rho_{eq}(x, \boldsymbol \lambda) \end{equation}
which follow from the definition of the nonequilibrium potential $ \phi $. These expressions may be used to rewrite Eq.~\eqref{eq:misc-compacttensor} as
\begin{align} \zeta_{ij}(\boldsymbol \lambda) = -\frac{Z(\boldsymbol \lambda)}{D} \int_{-\infty}^{\infty} dx ~  & \bigg[ \partial_{\lambda^{i}} \rho_{eq}(x, \boldsymbol \lambda)  \times \nonumber \\ &  \int_{a}^{x} dx' ~ e^{\beta U(x', \boldsymbol \lambda) } \partial_{\lambda^{j}} \Pi_{eq}(x', \boldsymbol \lambda) \bigg] \end{align}
or 
\begin{align} \zeta_{ij}(\boldsymbol \lambda) = -\frac{Z(\boldsymbol \lambda)}{D} \int_{-\infty}^{\infty} dx ~  & \bigg[ \partial_{x} \bigg( \partial_{\lambda^{i}} \Pi_{eq}(x, \boldsymbol \lambda) \bigg) \times \nonumber \\ &  \int_{a}^{x} dx' ~ e^{\beta U(x', \boldsymbol \lambda) } \partial_{\lambda^{j}} \Pi_{eq}(x', \boldsymbol \lambda) \bigg]. \end{align}
If 
\begin{equation}
\lim_{x \rightarrow \pm \infty} \partial_{\lambda^{i}} \Pi_{eq}(x, \boldsymbol \lambda)  \int_{a}^{x} dx' ~ e^{\beta U(x', \boldsymbol \lambda) } \partial_{\lambda^{j}} \Pi_{eq}(x', \boldsymbol \lambda) = 0,
\end{equation}
then

\begin{equation}\label{eq:misc-tensorcomps} \zeta_{ij}(\boldsymbol \lambda) = \frac{1}{D} \int_{-\infty}^{\infty} dx ~  \bigg[ \frac{\partial_{\lambda^{i}} \Pi_{eq}(x, \boldsymbol \lambda) ~\partial_{\lambda^{j}} \Pi_{eq}(x , \boldsymbol \lambda)}{\rho_{eq}(x,\boldsymbol \lambda)} \bigg],
 \end{equation}
which is our desired result.  Eq.~\eqref{eq:misc-tensorcomps} allows us to bypass the need for computing correlation functions in order to find the inverse diffusion tensor as it is based solely on equilibrium information. We will now explore an alternate approach to deriving this formula, and then we will apply it to two specific systems of interest.

\section{Derivative Truncation Approximation}
We relied on linear response theory to arrive at Eq.~\eqref{eq:zetaPi}. However, the inverse diffusion tensor arises naturally from a first order expansion in temporal derivatives of the control parameters as noted in~\cite{Zulkowski2012} for harmonic potentials. Assuming the probability distribution $ \rho(x,t) $ may be well approximated by
\begin{equation} \rho(x,t) \approx \rho_{eq}(x,\boldsymbol \lambda(t)) + \mathcal{G}(x,\lambda(t)) \cdot \frac{ d \boldsymbol \lambda}{dt} \end{equation}
where $ \mathcal{G} $ is determined by the Fokker-Planck equation,
we provide an alternative construction of the inverse diffusion tensor via the so-called ``derivative truncation" argument~\cite{Zulkowski2012}.

%Since we assume general, non-harmonic potentials, the moments appearing in the mean Y-value functional will not satisfy a closed, finite set of ordinary differential equations. Therefore, the construction used in~\cite{Zulkowski2012} does not obviously carry over to the more general situation. Instead, we base our proof on the mathematics developed in this chapter.

Referring to Eq.~\eqref{eq:FP}, we assume the nonequilibrium probability density has the approximate form
\begin{equation}\label{eq:misc-rhoapprox} \rho(x,t) \approx \rho_{eq}(x, \boldsymbol \lambda(t)) + \frac{d \lambda^{i}}{dt } \mathcal{G}_{i}(x, \boldsymbol \lambda(t)), \end{equation}
where $ \mathcal{G}_{i}(x, \boldsymbol \lambda) $ is to be determined. Substituting this expression into Eq.~\eqref{eq:FP} and neglecting higher-order derivatives, we see 
\begin{equation}\label{eq:misc-G} \frac{\partial \rho_{eq}(x, \boldsymbol \lambda)}{\partial \lambda^{i} } = D \bigg[ \frac{\partial }{\partial x} \bigg( \frac{\partial \left( \beta U(x, \boldsymbol \lambda) \right)}{\partial x} \mathcal{G}_{i}(x, \boldsymbol \lambda) \bigg) + \frac{\partial^2 }{\partial x^2} \mathcal{G}_{i}(x, \boldsymbol \lambda)  \bigg].
\end{equation}
Furthermore, since both $ \rho(x,t) $ and $ \rho_{eq}(x, \boldsymbol \lambda) $ are normalized probability distributions, we have the constraint\begin{equation} \label{eq:misc-Gint} \int_{-\infty}^{\infty} dx~\mathcal{G}_{i}(x , \boldsymbol \lambda) = 0. \end{equation}
We may systematically integrate Eq.~\eqref{eq:misc-G} to obtain a solution that also satisfies Eq.~\eqref{eq:misc-Gint} and appropriate boundary conditions. For our purposes, a more expedient way of arriving at the solution is to simply state a candidate and demonstrate that it satisfies the necessary requirements. 

Our candidate is
\begin{equation}\label{eq:misc-candidate} \mathcal{G}_{j}(x,\boldsymbol \lambda) \equiv -\langle \partial_{x}H_{2}(x; \boldsymbol \lambda, x_{0}) \partial_{\lambda^j} \phi(x_{0}, \boldsymbol \lambda) \rangle_{eq, \boldsymbol \lambda}. \end{equation}
Here, $ H_{2} $ is defined by Eqs.~\eqref{eq:misc-Hseq0},~\eqref{eq:misc-Hseq1} and~\eqref{eq:misc-Hseq2} and appropriate boundary conditions as dictated by the physics of the problem. The average $ \langle \cdot \rangle_{eq,\boldsymbol \lambda} $ applies to the variable $ x_{0} $ and is defined in terms of the stationary state probability distribution characterized by $ \boldsymbol \lambda $. 

We can quickly establish Eq.~\eqref{eq:misc-Gint}. Bringing the integral $ \int_{-\infty}^{\infty} dx $ inside of the stationary state average in Eq.~\eqref{eq:misc-candidate}, we see that $ \int_{-\infty}^{\infty} dx~\partial_{x} H_{2}(x; \boldsymbol \lambda , x_{0}) $  vanishes by the fundamental theorem of calculus in the case of an unbounded potential at $ \pm \infty $. 

Substituting Eq.~\eqref{eq:misc-candidate} into Eq.~\eqref{eq:misc-G}, we see that we must prove
%establish
\begin{align} \frac{\partial \rho_{eq}(x, \boldsymbol \lambda)}{\partial \lambda^{j} } = &-D \partial_{x} \big[ \big< \big( \beta U'(x, \boldsymbol \lambda) \partial_{x} H_{2}(x; \boldsymbol \lambda, x_{0}) \nonumber \\ &+ \partial_{x} H_{2}(x; \boldsymbol \lambda, x_{0}) \big) \partial_{\lambda^j} \phi(x_{0}, \boldsymbol \lambda) \big> _{eq, \boldsymbol \lambda} \big]. \end{align}
From Eq.~\eqref{eq:misc-Hseq2} follows
\begin{equation} D \big[ \beta U'(x, \boldsymbol \lambda) \partial_{x} H_{2}(x; \boldsymbol \lambda, x_{0}) + \partial_{x} H_{2}(x; \boldsymbol \lambda, x_{0}) \big] = H_{1}(x; \boldsymbol \lambda, x_{0}) \end{equation}
and so we must show
\begin{equation}  \frac{\partial \rho_{eq}(x, \boldsymbol \lambda)}{\partial \lambda^{j} } = -\langle \partial_{x}H_{1}(x; \boldsymbol \lambda, x_{0}) \partial_{\lambda^j} \phi(x_{0}, \boldsymbol \lambda) \rangle_{eq, \boldsymbol \lambda}. \end{equation}
For unbounded-potential boundary conditions, $ H_{1}(x;\boldsymbol \lambda, x_{0}) = \theta(x-x_{0}) + \Pi_{eq}(x,  \boldsymbol \lambda) $. Therefore,
\begin{align} \langle \partial_{x}H_{1}(x; \boldsymbol \lambda, x_{0}) \partial_{\lambda^j} \phi(x_{0}, \boldsymbol \lambda) \rangle_{eq, \boldsymbol \lambda}  = \langle \delta_{x,x_{0}} \partial_{\lambda^j} \phi(x_{0}, \boldsymbol \lambda) \rangle_{eq, \boldsymbol \lambda}. \end{align}
The second term vanishes since $ \rho_{eq}(x, \boldsymbol \lambda) $ is independent of $ x_{0} $ and $ \langle \partial_{\lambda^{j}} \phi(x_{0}, \boldsymbol \lambda) \rangle_{eq,\boldsymbol \lambda} = 0 $. Moreover,
\begin{equation} \langle \delta_{x,x_{0}} \partial_{\lambda^j} \phi(x_{0}, \boldsymbol \lambda) \rangle_{eq, \boldsymbol \lambda}  = \rho_{eq} \partial_{\lambda^j} \phi = - \partial_{\lambda^{j}} \rho_{eq}, \end{equation}
establishing the claim.

We are now in position to relate this derivative truncation approximation to the inverse diffusion tensor approximation. Recall that
\begin{equation} \langle Y \rangle_{\boldsymbol \Lambda} \equiv \int_{0}^{\tau} dt \ \bigg[ \frac{ d \boldsymbol \lambda^{T} }{dt} \bigg] \cdot \bigg< \frac{\partial \phi}{\partial \boldsymbol \lambda}\big(\boldsymbol \lambda(t)\big) \bigg>_{\boldsymbol \Lambda}, \end{equation}
where 
\begin{equation} \bigg< \frac{\partial \phi}{\partial \lambda^{i}}\big(\boldsymbol \lambda(t)\big) \bigg>_{\boldsymbol \Lambda} \equiv \int_{-\infty}^{\infty} dx~\rho(x,t) \frac{\partial \phi}{\partial \lambda^{i}}\big(x,\boldsymbol \lambda(t) \big).\end{equation}
Using the derivative truncation approximation,
\begin{widetext}
\begin{align} \bigg< \frac{\partial \phi}{\partial \lambda^{i}}\big(\boldsymbol \lambda(t)\big) \bigg>_{\boldsymbol \Lambda} &\approx \frac{d \lambda^{j}}{dt} \int_{-\infty}^{\infty} dx~ \mathcal{G}_{j}(x, \boldsymbol \lambda(t)) \frac{\partial \phi}{\partial \lambda^{i}}\big(x,\boldsymbol \lambda(t) \big) \nonumber \\
& = - \frac{d \lambda^{j}}{dt} \int_{-\infty}^{\infty} dx~\langle \partial_{x}H_{2}(x; \boldsymbol \lambda(t), x_{0})  \partial_{\lambda^j} \phi(x_{0}, \boldsymbol \lambda(t)) \rangle_{eq, \boldsymbol \lambda(t)} \frac{\partial \phi}{\partial \lambda^{i}}\big(x,\boldsymbol \lambda(t) \big) \nonumber \\
& = - \frac{d \lambda^{j}}{dt}\bigg<  \bigg[ \int_{-\infty}^{\infty} dx~\partial_{x}H_{2}(x; \boldsymbol \lambda(t), x_{0})  \frac{\partial \phi}{\partial \lambda^{i}}\big(x,\boldsymbol \lambda(t) \big) \bigg] \partial_{\lambda^j} \phi(x_{0}, \boldsymbol \lambda(t)) \bigg>_{eq, \boldsymbol \lambda(t)} \nonumber \\
& = \frac{d \lambda^{j}}{dt}\bigg<  \bigg[ \int_{0}^{\infty}dt'~ \int_{-\infty}^{\infty} dx~\rho(x,t'; x_{0})  \frac{\partial \phi}{\partial \lambda^{i}}\big(x;\boldsymbol \lambda(t) \big)  \bigg] \partial_{\lambda^j} \phi(x_{0}, \boldsymbol \lambda(t)) \bigg>_{eq, \boldsymbol \lambda(t)} \nonumber \\
& = \frac{d \lambda^{j}}{dt} \int_{0}^{\infty} dt'~ \langle \partial_{\lambda^{i}} \phi(t') \partial_{\lambda^{j}} \phi(0) \rangle_{eq,\lambda(t) } .
\end{align}
\end{widetext}
Therefore, we see that the derivative truncation approximation reproduces the inverse diffusion tensor for general potentials in the overdamped regime [compare with Eq.~\eqref{eq:zetacomps}].

We note that an expression similar to our Eq.~\eqref{eq:zetaPi} appears in~\cite{Szabo2011} in which the authors consider anisotropic Langevin dynamics in two dimensions. A separation of time scales gives rise to slow dynamics for one of the system coordinates which can be accurately modeled using a Markovian description with a position-dependent friction. This friction in turn is related to the time integral of an autocorrelation function similar to the one used here to define the inverse diffusion tensor. A possible direction for future work would be to investigate potential connections between the two results. 

\section{The Erasure Model}

We consider the following model to represent a single classical bit of information: an overdamped Brownian colloidal particle diffusing in a one-dimensional double-well potential in contact with a thermal bath of temperature $T$~\cite{Lutz2009,Lutz2012}.
The wells are initially separated by a potential barrier whose height is much larger than the energy scale $ \beta^{-1} \equiv k_{B} T $ set by thermal fluctuations, ensuring stability of memory. Explicitly, we may write the potential as
\begin{equation}\label{eq:erasurepot}
%U(x, \lambda) = \frac{\alpha}{\beta }  \big | \lambda - | x -1 +\lambda | \big | 
U(x,\lambda) \equiv -\frac{1}{\beta} \log \Bigg [ \frac{ \alpha e^{-\alpha \left( x- 1 + \lambda \right) }}{\left(1+ e^{-\alpha \left( x- 1 + \lambda \right) } \right)^2}+ \frac{ \alpha e^{\alpha \left( x- 1 \right) }}{\left(1+ e^{\alpha \left( x- 1 \right) } \right)^2} \Bigg ],
\end{equation}
where $ x $ is a dimensionless spatial coordinate and $ \alpha \gg 1 $. Initially, $ \lambda = 2 $ and there are two distinct wells and a central barrier with height governed by $ \alpha $. As $ \lambda $ decreases to $ 0 $, the barrier height diminishes and the left-hand well shifts to merge with the right-hand well. 

The system is prepared so that the particle has equal probability of being found in either well. This may be achieved, for example, by selecting the initial position of the particle to be at the midpoint of the potential barrier and waiting a sufficiently long relaxation period~\cite{Lutz2009}. After this relaxation period has elapsed the particle has equal probability of being located to the left or right of the origin.
If the particle is found to the left (right) of the origin, the memory value is defined to be $ 1 $ ($0$).

We are primarily interested in optimizing finite-time erasure efficiency over cyclic protocols for the classical single bit model described above. In~\cite{Zulkowski2014}, constraining the initial and final probability distributions forced the optimal protocols to have jump discontinuities at the end points. This was to be expected based on experience with optimization in the context of stochastic thermodynamics in general~\cite{Seifert2007,Seifert2008,Aurell2011,SeifertReview_2012,Then2008} and erasure efficiency in particular~\cite{Esposito2010,Aurell2012}. These jump discontinuities warrant caution when defining thermodynamic quantities such as the average dissipated heat 
~\cite{Then2008, Zulkowski2014}. 

When classical information is being erased, the difference in Shannon entropies of the final and initial probability distributions must satisfy $ \triangle S \equiv S_{f} - S_{i} < 0 $, which would allow us to define the erasure efficiency $ \epsilon \equiv -\triangle S / \left( k_{B} \langle \beta Q \rangle_{\boldsymbol \Lambda} \right) $ as the ratio of this decrease in Shannon entropy to the average heat $ \langle Q \rangle_{\boldsymbol \Lambda} $ released into the thermal bath~\cite{Diana2013,Plenio2001}. Taking $ k_{B} \equiv 1 $, we see that 
\begin{equation} \epsilon = \frac{1}{1+  \langle \triangle s^{\text{tot}} \rangle_{\boldsymbol \Lambda_{\text{cyc}}} / \left(- \triangle S \right)}, \end{equation}
where 
\begin{equation} \langle \triangle s^{\text{tot}} \rangle_{\boldsymbol \Lambda_{\text{cyc}}} = \langle \beta Q \rangle_{\boldsymbol \Lambda_{\text{cyc}}} + \triangle S  \end{equation}
is the total average entropy production.
Our goal will be to minimize the total average entropy over protocols while constraining $ \triangle S $. The constraint forces us to optimize over protocols with jump discontinuities at the endpoints.  

We may express the total average entropy in terms of the initial and final probability distributions as well as the average work done over the erasure stage of the cycle~\cite{SeifertReview_2012, Zulkowski2014}: 
\begin{align} 
\langle  \triangle s^{tot} \rangle_{\boldsymbol \Lambda_{\text{cyc}}} & = \langle  \beta W \rangle_{\boldsymbol \Lambda_{\text{erase}}}- \int_{\mathbb{R}} dx~\rho [ \beta U + \ln \rho ] \Big |_{0}^{\tau},
\end{align}
where $ \tau $ is the duration of the erasure stage. This follows from
\begin{equation} \langle  \beta W \rangle_{\boldsymbol \Lambda_{\text{reset}}} = \int_{\mathbb{R}} \rho(x,\tau) [ \beta U(x,0) - \beta U(x,\tau) ]. \end{equation}

Using the approximation Eq.~\eqref{eq:misc-rhoapprox}, we see that 
\begin{align} 
\langle  \beta W \rangle_{\boldsymbol \Lambda_{\text{erase}}} & \approx \int_{0}^{\tau} dt~\zeta(\lambda(t)) \left( \frac{d \lambda}{dt} \right)^2- \ln \left( \frac{ Z_{\tau}}{Z_0} \right).
\end{align}  

During the first (erasure) stage, the initial equilibrium distribution transitions to a final nonequilibrium distribution in which the system is overwhelmingly more likely to have memory value $0$. In the second (reset) stage, $ \lambda $ is brought instantaneously back to its original value while keeping the particle probability distribution constant. No heat is generated during this stage.

Using the calculus of variations, we seek the twice-differentiable optimizer of $  \int_{0}^{\tau} dt~\zeta(\lambda(t)) \left(\frac{d \lambda}{dt} \right)^2 $ for the erasure stage of the cycle. By considering variations that vanish away from the endpoints of the protocol, this optimizer satisfies the Euler-Lagrange equation 
\begin{equation} \label{eq:optlambda}
\frac{d \lambda}{dt} = - \frac{ \int_{\lambda_{f}}^{\lambda_{i}} dz~\sqrt{\zeta(z)} \big / \tau}{\sqrt{\zeta(\lambda(t))}},
\end{equation}
where $ \lambda_{i} $ and $ \lambda_{f} $ are determined by the constraints on the probability distribution.

By definition, 
\begin{equation} \triangle S = - \int_{\mathbb{R}} dx~\rho(x,\tau) \ln \rho(x,\tau) + \int_{\mathbb{R}} dx~\rho(x,0) \ln \rho(x,0). \end{equation} 
Using Eq.~\eqref{eq:misc-rhoapprox}, we can derive an approximate expression for $ \triangle S $, which allows us to compute the endpoints of the protocol.
The average total entropy generated during the optimal protocol is approximately 
\begin{align} 
\langle & \triangle s^{tot} \rangle_{\boldsymbol \Lambda_{opt}}   \approx \frac{\left( \int_{\lambda_{i}}^{\lambda_{f}} dz~\sqrt{\zeta(z)}\right)^2}{\tau}+ \mathcal{O}(1/\tau^2).
\end{align}  
Since we are neglecting terms of order $ 1/\tau^2 $ and higher, we need only compute $ \lambda_{i} $ and $ \lambda_{f} $ assuming $ \rho(x,t) \approx \rho_{eq}(x,\lambda(t)) $ throughout the driving process. For simplicity, we assume that $ \lambda_{i} = 2 $ , $ \lambda_{f}= 0 $. 

From Eq.~\eqref{eq:misc-tensorcomps}, we have
\begin{equation} \label{eq:erasurezeta} \zeta(\lambda) = \frac{1}{2D} \Bigg \{ 1- \frac{\frac{\pi}{2}-\tan^{-1} \left( \sqrt{2 \big / \big [ \cosh( \alpha \lambda) -1 \big] } \right)}{\sqrt{ 2 \big [ \cosh( \alpha \lambda) -1 \big] }} \Bigg \}. \end{equation}
\begin{figure}[ht!]
\begin{center}
\includegraphics{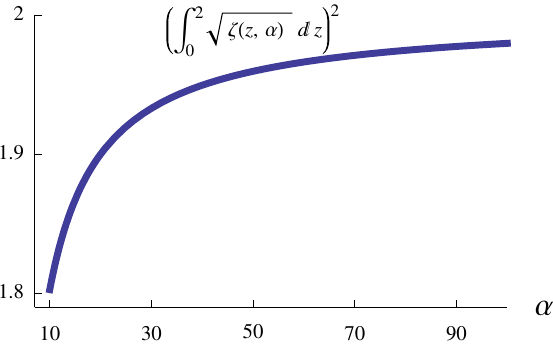}
\end{center}
\caption{Dependence of the leading order term of $ \langle \triangle s^{\text{tot}} \rangle_{\boldsymbol \Lambda_{\text{opt}}} $ on $ \alpha $ for the erasure model given by Eq.~\eqref{eq:erasurepot}. Note the saturation at $ 2 $ as $ \alpha $ becomes very large. }
\label{fig:alphadep}
\end{figure}

The dependence of the leading order term of the optimal total average entropy on $ \alpha $ is displayed in Fig.~\ref{fig:alphadep}. We note that the leading order term grows as a function of $ \alpha $ and appears to approach a limiting value close to $ 2 $ as $ \alpha $ becomes very large. In~\cite{Zulkowski2014}, a similar erasure model is considered. There, the potential is piecewise constant with central barrier height and ``tilt" of the lefthand well as control parameters. The leading order term of $ \langle \triangle s^{\text{tot}} \rangle_{\boldsymbol \Lambda_{\text{opt}}} $ for that model is given in terms of the Hellinger distance between the initial and final probability distributions:
\begin{equation}   
4 \left [ \sqrt{p_{r}(\tau)} - \sqrt{p_{r}(0)} \right ]^2 + 4 \left [ \sqrt{p_{l}(\tau)} - \sqrt{p_{l}(0)} \right ]^2.  
\end{equation}
Assuming the particle has overwhelming probability to be located in the righthand well after the erasure cycle, we see that $ \tau \langle \triangle s^{\text{tot}} \rangle_{\boldsymbol \Lambda_{\text{opt}}} \approx 2.343 $ for barrier heights of physical relevance. (Compare with the current model's saturation value of $ \tau \langle \triangle s^{\text{tot}} \rangle_{\boldsymbol \Lambda_{\text{opt}}} \approx 2 $.) 
%A potential future research avenue would address the apparent
In future work, it will be interesting to explore whether there might be a common cause for the intriguing similarity between the saturation values for the leading order terms in the optimal average entropy production for these two different erasure cycle models.

\section{The Ratchet model}

We consider an overdamped colloidal particle diffusing in one dimension subject to optical trap confinement and a tilted ratchet potential~\cite{Reimann2002}. Mathematically, 
\begin{equation}\label{eq:misc-ratchetpotential} \beta U(x,  \boldsymbol \lambda) = \frac{1}{2} \beta k \left( x-x_{0} \right)^2 - \beta F x + \beta V_{0} ~\varphi \left( x / l \right). \end{equation} 
Here, $ l $ is some characteristic length scale, $ F $ is the ``tilt" of the ratchet, and $ V_{0} $ is the magnitude of the ratchet potential. Furthermore, we choose a single control parameter, namely $ x_{0} $, the position of the center of the optical trap.

The model possesses relative mathematical simplicity and captures essential physics of chemical processes relevant to the operation of cellular machinery. We view the following as the first step towards applying the inverse diffusion tensor framework to optimization of nonequilibrium processes underlying the functionality of molecular motors and other nanoscale biological machines.

If we consider an isothermal chemical reaction in the presence of a catalyst protein (\emph{i.e.} an enzyme), then the reaction can be described by a single reaction coordinate, cycling through a number of chemical states in the simplest case~\cite{Reimann2002}. A suitable working model is then an overdamped Brownian particle (reaction coordinate) in the presence of thermal fluctuations in a periodic potential. For this reason, we select $ \varphi(y) = \sin(y) $. 

If the concentrations of the reactants and products are away from their equilibrium ratio, then the catalyst molecule will loop through the chemical reaction cycle preferably in one direction~\cite{Reimann2002}. In the corresponding ratchet model, the periodic potential must be supplemented by a constant tilt, \emph{i.e.} $ F$. 

Define
\begin{align} \alpha &\equiv \beta k l^2 \ , \ y \equiv x/l \ , \ f \equiv F / (k l) \ , \ \nonumber \\  u_{0}(y, y_{0}) & \equiv \frac{1}{2}(y-y_{0})^2 - f y  \ , \  \epsilon \equiv V_{0}/ (k l^2) \end{align}
so that $ \beta U(x, x_{0}) = \alpha \big[ u_{0}(y, y_{0}) + \epsilon \varphi(y) \big] $. Assuming the strength of the optical trap far exceeds the strength of the ratchet potential, $ \epsilon $ is a small parameter and we may apply perturbation theory; \emph{i.e.} we expand all quantities to first order in $ \epsilon $ and discard higher order terms.

It is straightforward to show that
\begin{equation} \label{eq:ratcheteqdist}
\rho_{eq}(y,y_{0}) \approx \frac{1}{l} \sqrt{\frac{\alpha}{2 \pi}} e^{-\frac{1}{2} \alpha ( y - y_{0}-f )^2} \bigg[ 1 - \alpha \epsilon \big( \varphi(y)- \langle \varphi \rangle_{0} \big) \bigg] 
\end{equation}
where the subscript ``0" indicates an average with $ \epsilon = 0$. 

Using Eq.~\eqref{eq:misc-tensorcomps}, we compute
\begin{equation}
\zeta(y_{0}) \approx \frac{l^2}{D}  \left( 1- 2 \alpha \epsilon \langle \varphi'' \rangle_{0} \right). 
\end{equation}
The equilibrium distribution for the ratchet model (Eq.~\eqref{eq:ratcheteqdist}) 
%is 
%\begin{equation} \label{eq:appprob}
%\rho_{eq}(y,y_{0}) \approx \frac{1}{l} \sqrt{\frac{\alpha}{2 \pi}} e^{-\frac{1}{2} \alpha ( y - y_{0}-f )^2} \bigg[ 1 - \alpha \epsilon \big( \varphi(y)- \langle \varphi \rangle_{0} \big) \bigg]  
%\end{equation}
has cumulative distribution function 
\begin{equation} \label{eq:cumdist}
\Pi_{eq}(y,y_{0}) = \int_{-\infty}^{y} dy' \rho_{eq}(y',y_{0}). 
\end{equation}
%Since
%\begin{equation}
%\partial_{y_{0}} \langle \varphi \rangle_{0} = \langle \varphi' \rangle_{0} ,
%\end{equation}
%we have that
%\begin{widetext}
%\begin{align}\label{eq:appcum2}
%& \partial_{y_{0}} \Pi_{eq}(y,y_0)  = \sqrt{\frac{\alpha}{2\pi}} \bigg \{ -e^{-\frac{1}{2} \alpha \left( y-y_{0}-f \right)^2} + \alpha \epsilon \bigg( \big( \varphi(y)- \langle \varphi \rangle_{0} \big) e^{-\frac{1}{2} \alpha \left( y-y_{0}-f \right)^2} - \int_{-\infty}^{y} dy' e^{-\frac{1}{2} \alpha \left( y'-y_{0}-f \right)^2} \left( \varphi'(y')-\langle \varphi' \rangle_{0} \right) \bigg) \bigg \} \nonumber \\
% & \partial_{y_{0}}  \Pi_{eq}(y,y_0)~e^{\beta  U(y,y_{0}) }  =  \sqrt{\frac{\alpha}{2\pi}} e^{-\alpha \left( f y_{0} + \frac{1}{2} f^2 \right)} \bigg \{ -1 +  \alpha \epsilon \bigg( \varphi(y)- \langle \varphi \rangle_{0}  -  e^{\frac{1}{2} \alpha \left( y-y_{0}-f \right)^2}  \int_{y}^{\infty} dy' e^{-\frac{1}{2} \alpha \left( y'-y_{0}-f \right)^2}  \left( \varphi'(y')-\langle \varphi' \rangle_{0} \right) \bigg) \bigg \}. 
%\end{align}
%\end{widetext}
According to~\cite{Stegun1972}, 
\begin{equation} 
e^{w^2} \int_{w}^{\infty} e^{-z^2} dz \leq \frac{1}{w +\sqrt{w^2+\frac{4}{\pi}}} 
\end{equation}
for $ w \geq 0 $. Since $ \varphi $ and all of its derivatives are bounded on $ \mathbb{R} $, 
\begin{equation}
\lim_{y \rightarrow \infty} \partial_{y_{0}} \Pi_{eq}(y, y_{0})  \int_{0}^{y} dy' ~ e^{\beta U(y', y_{0}) } \partial_{y_{0}} \Pi_{eq}(y', y_0) = 0.
\end{equation}
A similar argument shows that 
\begin{equation}
\lim_{y \rightarrow -\infty} \partial_{y_{0}} \Pi_{eq}(y, y_{0})  \int_{0}^{y} dy' ~ e^{\beta U(y', y_{0}) } \partial_{y_{0}} \Pi_{eq}(y', y_0) = 0,
\end{equation}
and so we may apply Eq.~\eqref{eq:misc-tensorcomps} to compute the inverse diffusion tensor.

Using Eqs.~\eqref{eq:ratcheteqdist} and~\eqref{eq:cumdist}, we see that
\begin{widetext}
\begin{align}
 \frac{\big [ \partial_{y_{0}} \Pi_{eq} \big ]^2 }{\rho_{eq} }  \approx l \sqrt{\frac{\alpha}{2\pi}} \bigg \{ e^{-\frac{\alpha}{2} \left( y-y_0-f \right)^2} + \alpha \epsilon \bigg( -( \varphi(y)- \langle \varphi \rangle_{0} )  e^{-\frac{\alpha}{2} \left( y-y_0-f \right)^2} + 2 \int_{-\infty}^{y} dy' \left( \varphi'(y')-\langle \varphi' \rangle_{0} \right) e^{-\frac{\alpha}{2} \left( y'-y_0-f \right)^2} \bigg) \bigg \}.
\end{align}
\end{widetext}
From Eq.~\eqref{eq:misc-tensorcomps} we have
\begin{align}
\zeta(y_{0})  \approx \frac{l^2}{D} \bigg \{ &1+ 2 \alpha \epsilon \int_{-\infty}^{\infty} dy \int_{-\infty}^{y} dy' ~ \left( \varphi'(y')- \langle \varphi' \rangle_0 \right) \times \nonumber \\ & e^{-\frac{\alpha}{2} \left( y'-y_0-f \right)^2}  \bigg / \sqrt{\frac{2 \pi}{\alpha}} \bigg \}.
\end{align}
The integral may be evaluated using integration by parts. Therefore, we have
%\begin{widetext}
%\begin{align}
%& \int_{-\infty}^{\infty} dy \int_{-\infty}^{y} dy' ~ \left( \varphi'(y')- \langle \varphi' \rangle_0 \right) e^{-\frac{\alpha}{2} \left( y'-y_0-f \right)^2}  = \int_{-\infty}^{\infty} dy \bigg \{\frac{d}{dy} \left( y-y_{0}-f \right)  \int_{-\infty}^{y} dy'  \left( \varphi'(y')- \langle \varphi' \rangle_0 \right) e^{-\frac{\alpha}{2} \left( y'-y_0-f \right)^2} \bigg \} \nonumber \\
%&= \int_{-\infty}^{\infty} dy \left( y-y_{0}-f \right) \left( \langle \varphi' \rangle_0 - \varphi'(y) \right) e^{-\frac{\alpha}{2} \left( y-y_0-f \right)^2} = 
%\int_{-\infty}^{\infty} dy \left( \varphi'(y)- \langle \varphi' \rangle_0 \right) \frac{d}{dy} \left(  \frac{1}{\alpha} e^{-\frac{\alpha}{2} \left( y-y_0-f \right)^2} \right) = \nonumber \\
%&- \frac{1}{\alpha} \int_{-\infty}^{\infty} dy~\varphi''(y)  e^{-\frac{\alpha}{2} \left( y-y_0-f \right)^2} = -\sqrt{\frac{2 \pi}{\alpha}} \langle \varphi'' \rangle_{0}.
%\end{align}
%\end{widetext}
\begin{equation}
\zeta(y_{0}) \approx \frac{l^2}{D} \left( 1- 2 \alpha \epsilon \langle \varphi'' \rangle_{0} \right). 
\end{equation}

For $ \varphi(y) = \sin(y) $ (which is mathematically similar to the parametric quantron of~\cite{Likharev1982}),
\begin{equation}
\zeta(y_{0}) \approx \frac{l^2}{D}  \left( 1 + 2 \epsilon \ e^{-\frac{1}{2 \alpha}} \sin \left( y_{0}+ f \right)  \right). 
\end{equation}

\begin{figure}
\begin{center}
\includegraphics{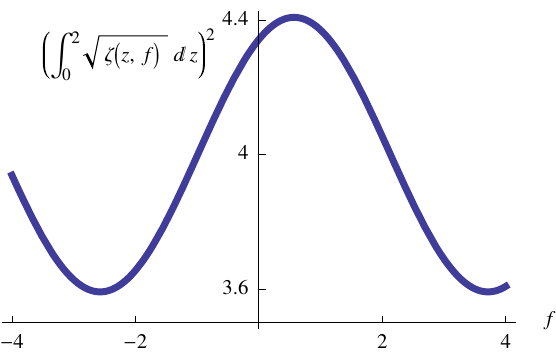}
\caption{Dependence of minimal dissipation for ratchet model on (scaled) tilt $ f $ where $ \alpha = 1 $,  $ \epsilon = 0.1 $ and $ y_{0}(0) = 0 $ and $ y_{0}(\tau) = 2 $. Note that the minimal dissipation is a periodic function of $ f $.}
\label{fig:ratchet}
\end{center}
\end{figure}

Since
\begin{equation} \langle \beta W_{ex} \rangle_{\boldsymbol \Lambda} \approx \int_{0}^{\tau} dt~\zeta(y_{0}) \left( \frac{d y_{0}}{dt} \right)^2, \end{equation}
the minimal dissipation is
\begin{equation} \langle \beta W_{ex} \rangle_{\boldsymbol \Lambda_{\text{opt}}} \approx \frac{\left( \int_{y_{0}(0)}^{y_{0}(\tau)} dz~\sqrt{\zeta(z)}\right)^2}{\tau}.\end{equation}
We plot the dependence of this expression on $ f $ in Fig.~\ref{fig:ratchet}  which clearly illustrates the periodicity in the tilt. This periodicity has a natural interpretation. A shift in the tilt $ f $ by $ 2 \pi $ units corresponds to an overall shift in the potential energy by a constant amount and a horizontal translation upon completing the square. Since constant shifts in the potential are physically irrelevant, we would expect the optimal average excess work to be insensitive to such a shift in the tilt. 
%the optimal time course for the trap center is given by the Euler-Lagrange equation
%\begin{equation} \frac{d y_{0}}{dt} = \frac{ C}{ \sqrt{ \frac{ D}{l^2}} \sqrt{\zeta(y_{0})}}. \end{equation}
%Here, $ C $ is a constant of integration.
%
%Using perturbation theory, an approximate solution is given by
%\begin{equation}
%y_{0}(t) \approx C t + B + \epsilon~e^{-\frac{1}{2 \alpha}} \cos \left( C t + f + A \right) 
%\end{equation}
%where the constants $ A,B$ and $ C $ are determined by the endpoints of the protocol and the duration $ \tau $. Note that the temporal periodicity of the optimal protocol driving the center of the harmonic trap reflects the spatial periodicity of the ratchet.

\section{Discussion}
In this paper, we constructed a compact formula (Eq.~\eqref{eq:zetaPi}) for the inverse diffusion tensor for a broad class of overdamped dynamics. This is a powerful expression in part because the inverse diffusion tensor components are expressed entirely in terms of the equilibrium probability distribution, and we expect it to increase the number of applications for this approach.

We applied this formula to calculate the optimal dissipation for two model systems. In the first model, we considered the erasure of a single classical bit of information. The system modeling the storage and erasure of this bit consisted of a colloidal particle diffusing under Brownian dynamics in a double-well potential. In the second model, we considered a colloidal Brownian particle coupled to an optical trap and a tilted ratchet potential.
In both cases, the inverse diffusion tensor allowed us to predict the optimal time course and produced an explicit expression for the minimal dissipation.

Consistent with previous studies~\cite{Esposito2010,Aurell2012,Diana2013,Zulkowski2014}, we find that the minimal dissipation for each model considered here obeys the expected $ 1 / \tau $ law. As with the erasure model we considered previously~\cite{Zulkowski2014}, the coefficient multiplying the $1/\tau$ term is fixed by the diffusion tensor, but it does not obviously take the same functional form of the square of the Hellinger distance between initial and final probability distributions.  A direct comparison between the two erasure models is challenging given that the previous example was a discrete three state system whereas here our system is continuous. Yet, despite these differences,
% \prz{In comparing the leading order term of the optimal average entropy for the erasure model considered here with the corresponding quantity for the erasure model studied in~\cite{Zulkowski2014}, 
we found a striking similarity between the saturation values for large, physically relevant barrier heights. It is possible that the saturation value for the optimal average entropy produced during an erasure cycle possesses some universality as a function of initial barrier height.
Future work will address this intriguing possibility. 

Analyzing these erasure models in the regime of low to moderate initial barrier height provides another interesting avenue for future development. As described in~\cite{Pankratov1997}, characteristic time scales of dynamical systems driven by noise --- such as the relaxation time to a steady state --- follow from the moments of the transition time if a specific potential (\emph{e.g.}, Eq.~\eqref{eq:erasurepot}) is assumed. It may be possible to apply the framework of~\cite{Pankratov1997,Nikitenkova1998} to these erasure models especially in light of the fact that the inverse diffusion tensor may be used to compute the integral relaxation time for a broad class of systems driven by noise~\cite{Sivak2012}.

We expect further study of these models to generate yet more testable predictions for experiments in erasure similar to the setup of~\cite{Jun2014} and the parametric quantron based on the Josephson effect in superconductors as in~\cite{Likharev1982}. 
%\mrd{--- an exciting direction for future work.}
%, it will be interesting to explore whether there might be a common cause for the intriguing similarity between the saturation values for the leading order terms in the optimal average entropy production for these two different erasure cycle models.}

\acknowledgments
The authors would like to thank Dibyendu Mandal, David Sivak and Gavin Crooks for many helpful discussions.
M.\ R.\ D.\ gratefully acknowledges support from the McKnight Foundation and the Hellman Family Faculty Fund. M.\ R.\ D.\ and P.\ R.\ Z.\ were partly supported by the National Science Foundation through Grant No. IIS-1219199. 
This material is
based upon work supported in part by the US Army Research
Laboratory and the US Army Research Office under Contract
No. W911NF-13-1-0390.

\bibliography{generalpotentials_iter3}

\begin{thebibliography}{64}%
\makeatletter
\providecommand \@ifxundefined [1]{%
 \@ifx{#1\undefined}
}%
\providecommand \@ifnum [1]{%
 \ifnum #1\expandafter \@firstoftwo
 \else \expandafter \@secondoftwo
 \fi
}%
\providecommand \@ifx [1]{%
 \ifx #1\expandafter \@firstoftwo
 \else \expandafter \@secondoftwo
 \fi
}%
\providecommand \natexlab [1]{#1}%
\providecommand \enquote  [1]{``#1''}%
\providecommand \bibnamefont  [1]{#1}%
\providecommand \bibfnamefont [1]{#1}%
\providecommand \citenamefont [1]{#1}%
\providecommand \href@noop [0]{\@secondoftwo}%
\providecommand \href [0]{\begingroup \@sanitize@url \@href}%
\providecommand \@href[1]{\@@startlink{#1}\@@href}%
\providecommand \@@href[1]{\endgroup#1\@@endlink}%
\providecommand \@sanitize@url [0]{\catcode `\\12\catcode `\$12\catcode
  `\&12\catcode `\#12\catcode `\^12\catcode `\_12\catcode `\%12\relax}%
\providecommand \@@startlink[1]{}%
\providecommand \@@endlink[0]{}%
\providecommand \url  [0]{\begingroup\@sanitize@url \@url }%
\providecommand \@url [1]{\endgroup\@href {#1}{\urlprefix }}%
\providecommand \urlprefix  [0]{URL }%
\providecommand \Eprint [0]{\href }%
\providecommand \doibase [0]{http://dx.doi.org/}%
\providecommand \selectlanguage [0]{\@gobble}%
\providecommand \bibinfo  [0]{\@secondoftwo}%
\providecommand \bibfield  [0]{\@secondoftwo}%
\providecommand \translation [1]{[#1]}%
\providecommand \BibitemOpen [0]{}%
\providecommand \bibitemStop [0]{}%
\providecommand \bibitemNoStop [0]{.\EOS\space}%
\providecommand \EOS [0]{\spacefactor3000\relax}%
\providecommand \BibitemShut  [1]{\csname bibitem#1\endcsname}%
\let\auto@bib@innerbib\@empty
%</preamble>
\bibitem [{\citenamefont {Evans}\ \emph {et~al.}(1993)\citenamefont {Evans},
  \citenamefont {Cohen},\ and\ \citenamefont {Morriss}}]{Evans1993}%
  \BibitemOpen
  \bibfield  {author} {\bibinfo {author} {\bibfnamefont {D.~J.}\ \bibnamefont
  {Evans}}, \bibinfo {author} {\bibfnamefont {E.~G.~D.}\ \bibnamefont {Cohen}},
  \ and\ \bibinfo {author} {\bibfnamefont {G.~P.}\ \bibnamefont {Morriss}},\
  }\href@noop {} {\bibfield  {journal} {\bibinfo  {journal} {Phys. Rev. Lett.}\
  }\textbf {\bibinfo {volume} {71}},\ \bibinfo {pages} {2401} (\bibinfo {year}
  {1993})}\BibitemShut {NoStop}%
\bibitem [{\citenamefont {Evans}\ and\ \citenamefont
  {Searles}(1994)}]{Evans1994}%
  \BibitemOpen
  \bibfield  {author} {\bibinfo {author} {\bibfnamefont {D.~J.}\ \bibnamefont
  {Evans}}\ and\ \bibinfo {author} {\bibfnamefont {D.~J.}\ \bibnamefont
  {Searles}},\ }\href@noop {} {\bibfield  {journal} {\bibinfo  {journal} {Phys.
  Rev. E}\ }\textbf {\bibinfo {volume} {50}},\ \bibinfo {pages} {1645}
  (\bibinfo {year} {1994})}\BibitemShut {NoStop}%
\bibitem [{\citenamefont {Gallavotti}\ and\ \citenamefont
  {Cohen}(1995)}]{Gallavotti1995a}%
  \BibitemOpen
  \bibfield  {author} {\bibinfo {author} {\bibfnamefont {G.}~\bibnamefont
  {Gallavotti}}\ and\ \bibinfo {author} {\bibfnamefont {E.~G.~D.}\ \bibnamefont
  {Cohen}},\ }\href@noop {} {\bibfield  {journal} {\bibinfo  {journal} {Phys.
  Rev. Lett.}\ }\textbf {\bibinfo {volume} {74}},\ \bibinfo {pages} {2694}
  (\bibinfo {year} {1995})}\BibitemShut {NoStop}%
\bibitem [{\citenamefont {Crooks}(1999)}]{Crooks1999}%
  \BibitemOpen
  \bibfield  {author} {\bibinfo {author} {\bibfnamefont {G.~E.}\ \bibnamefont
  {Crooks}},\ }\href@noop {} {\bibfield  {journal} {\bibinfo  {journal} {Phys.
  Rev. E}\ }\textbf {\bibinfo {volume} {60}},\ \bibinfo {pages} {2721}
  (\bibinfo {year} {1999})}\BibitemShut {NoStop}%
\bibitem [{\citenamefont {Hatano}\ and\ \citenamefont
  {Sasa}(2001)}]{Hatano2001}%
  \BibitemOpen
  \bibfield  {author} {\bibinfo {author} {\bibfnamefont {T.}~\bibnamefont
  {Hatano}}\ and\ \bibinfo {author} {\bibfnamefont {S.}~\bibnamefont {Sasa}},\
  }\href@noop {} {\bibfield  {journal} {\bibinfo  {journal} {Phys. Rev. Lett.}\
  }\textbf {\bibinfo {volume} {86}},\ \bibinfo {pages} {3463} (\bibinfo {year}
  {2001})}\BibitemShut {NoStop}%
\bibitem [{\citenamefont {Wang}\ \emph {et~al.}(2002)\citenamefont {Wang},
  \citenamefont {Sevick}, \citenamefont {Mittag}, \citenamefont {Searles},\
  and\ \citenamefont {Evans}}]{Wang2002}%
  \BibitemOpen
  \bibfield  {author} {\bibinfo {author} {\bibfnamefont {G.~M.}\ \bibnamefont
  {Wang}}, \bibinfo {author} {\bibfnamefont {E.~M.}\ \bibnamefont {Sevick}},
  \bibinfo {author} {\bibfnamefont {E.}~\bibnamefont {Mittag}}, \bibinfo
  {author} {\bibfnamefont {D.~J.}\ \bibnamefont {Searles}}, \ and\ \bibinfo
  {author} {\bibfnamefont {D.~J.}\ \bibnamefont {Evans}},\ }\href@noop {}
  {\bibfield  {journal} {\bibinfo  {journal} {Phys. Rev. Lett.}\ }\textbf
  {\bibinfo {volume} {89}},\ \bibinfo {pages} {050601} (\bibinfo {year}
  {2002})}\BibitemShut {NoStop}%
\bibitem [{\citenamefont {Carberry}\ \emph {et~al.}(2004)\citenamefont
  {Carberry}, \citenamefont {Reid}, \citenamefont {Wang}, \citenamefont
  {Sevick}, \citenamefont {Searles},\ and\ \citenamefont
  {Evans}}]{Carberry2004}%
  \BibitemOpen
  \bibfield  {author} {\bibinfo {author} {\bibfnamefont {D.~M.}\ \bibnamefont
  {Carberry}}, \bibinfo {author} {\bibfnamefont {J.~C.}\ \bibnamefont {Reid}},
  \bibinfo {author} {\bibfnamefont {G.~M.}\ \bibnamefont {Wang}}, \bibinfo
  {author} {\bibfnamefont {E.~M.}\ \bibnamefont {Sevick}}, \bibinfo {author}
  {\bibfnamefont {D.~J.}\ \bibnamefont {Searles}}, \ and\ \bibinfo {author}
  {\bibfnamefont {D.~J.}\ \bibnamefont {Evans}},\ }\href@noop {} {\bibfield
  {journal} {\bibinfo  {journal} {Phys. Rev. Lett.}\ }\textbf {\bibinfo
  {volume} {92}},\ \bibinfo {pages} {140601} (\bibinfo {year}
  {2004})}\BibitemShut {NoStop}%
\bibitem [{\citenamefont {Garnier}\ and\ \citenamefont
  {Ciliberto}(2005)}]{Garnier2005}%
  \BibitemOpen
  \bibfield  {author} {\bibinfo {author} {\bibfnamefont {N.}~\bibnamefont
  {Garnier}}\ and\ \bibinfo {author} {\bibfnamefont {S.}~\bibnamefont
  {Ciliberto}},\ }\href@noop {} {\bibfield  {journal} {\bibinfo  {journal}
  {Phys. Rev. E}\ }\textbf {\bibinfo {volume} {71}},\ \bibinfo {pages} {060101}
  (\bibinfo {year} {2005})}\BibitemShut {NoStop}%
\bibitem [{\citenamefont {Toyabe}\ \emph {et~al.}(2010)\citenamefont {Toyabe},
  \citenamefont {Sagawa}, \citenamefont {Ueda}, \citenamefont {Muneyuki},\ and\
  \citenamefont {Sano}}]{Toyabe2010}%
  \BibitemOpen
  \bibfield  {author} {\bibinfo {author} {\bibfnamefont {S.}~\bibnamefont
  {Toyabe}}, \bibinfo {author} {\bibfnamefont {T.}~\bibnamefont {Sagawa}},
  \bibinfo {author} {\bibfnamefont {M.}~\bibnamefont {Ueda}}, \bibinfo {author}
  {\bibfnamefont {E.}~\bibnamefont {Muneyuki}}, \ and\ \bibinfo {author}
  {\bibfnamefont {M.}~\bibnamefont {Sano}},\ }\href@noop {} {\bibfield
  {journal} {\bibinfo  {journal} {Nat. Phys.}\ }\textbf {\bibinfo {volume}
  {6}},\ \bibinfo {pages} {988} (\bibinfo {year} {2010})}\BibitemShut {NoStop}%
\bibitem [{\citenamefont {Szilard}(1929)}]{Szilard1929}%
  \BibitemOpen
  \bibfield  {author} {\bibinfo {author} {\bibfnamefont {L.}~\bibnamefont
  {Szilard}},\ }\href@noop {} {\bibfield  {journal} {\bibinfo  {journal}
  {Z.Phys.}\ }\textbf {\bibinfo {volume} {53}},\ \bibinfo {pages} {840}
  (\bibinfo {year} {1929})}\BibitemShut {NoStop}%
\bibitem [{\citenamefont {Landauer}(1961)}]{Landauer1961}%
  \BibitemOpen
  \bibfield  {author} {\bibinfo {author} {\bibfnamefont {R.}~\bibnamefont
  {Landauer}},\ }\href@noop {} {\bibfield  {journal} {\bibinfo  {journal} {IBM
  J. Res. Dev.}\ }\textbf {\bibinfo {volume} {5}},\ \bibinfo {pages} {183}
  (\bibinfo {year} {1961})}\BibitemShut {NoStop}%
\bibitem [{\citenamefont {Bennett}(1982)}]{Bennett1982}%
  \BibitemOpen
  \bibfield  {author} {\bibinfo {author} {\bibfnamefont {C.}~\bibnamefont
  {Bennett}},\ }\href@noop {} {\bibfield  {journal} {\bibinfo  {journal} {Int.
  J. Theor. Phys.}\ }\textbf {\bibinfo {volume} {21}},\ \bibinfo {pages} {905}
  (\bibinfo {year} {1982})}\BibitemShut {NoStop}%
\bibitem [{\citenamefont {Jarzynski}(1997)}]{Jarzynski1997}%
  \BibitemOpen
  \bibfield  {author} {\bibinfo {author} {\bibfnamefont {C.}~\bibnamefont
  {Jarzynski}},\ }\href@noop {} {\bibfield  {journal} {\bibinfo  {journal}
  {Phys. Rev. Lett.}\ }\textbf {\bibinfo {volume} {78}},\ \bibinfo {pages}
  {2690} (\bibinfo {year} {1997})}\BibitemShut {NoStop}%
\bibitem [{\citenamefont {Liphardt}\ \emph {et~al.}(2002)\citenamefont
  {Liphardt}, \citenamefont {Dumont}, \citenamefont {Smith}, \citenamefont
  {Tinoco~Jr},\ and\ \citenamefont {Bustamante}}]{Liphardt2002}%
  \BibitemOpen
  \bibfield  {author} {\bibinfo {author} {\bibfnamefont {J.~T.}\ \bibnamefont
  {Liphardt}}, \bibinfo {author} {\bibfnamefont {S.}~\bibnamefont {Dumont}},
  \bibinfo {author} {\bibfnamefont {S.~B.}\ \bibnamefont {Smith}}, \bibinfo
  {author} {\bibfnamefont {I.}~\bibnamefont {Tinoco~Jr}}, \ and\ \bibinfo
  {author} {\bibfnamefont {C.}~\bibnamefont {Bustamante}},\ }\href@noop {}
  {\bibfield  {journal} {\bibinfo  {journal} {Science}\ }\textbf {\bibinfo
  {volume} {296}},\ \bibinfo {pages} {1832} (\bibinfo {year}
  {2002})}\BibitemShut {NoStop}%
\bibitem [{\citenamefont {Seifert}(2005)}]{Seifert2005b}%
  \BibitemOpen
  \bibfield  {author} {\bibinfo {author} {\bibfnamefont {U.}~\bibnamefont
  {Seifert}},\ }\href@noop {} {\bibfield  {journal} {\bibinfo  {journal} {Phys.
  Rev. Lett.}\ }\textbf {\bibinfo {volume} {95}},\ \bibinfo {pages} {040602}
  (\bibinfo {year} {2005})}\BibitemShut {NoStop}%
\bibitem [{\citenamefont {Sagawa}\ and\ \citenamefont
  {Ueda}(2010)}]{Sagawa2010}%
  \BibitemOpen
  \bibfield  {author} {\bibinfo {author} {\bibfnamefont {T.}~\bibnamefont
  {Sagawa}}\ and\ \bibinfo {author} {\bibfnamefont {M.}~\bibnamefont {Ueda}},\
  }\href@noop {} {\bibfield  {journal} {\bibinfo  {journal} {Phys. Rev. Lett.}\
  }\textbf {\bibinfo {volume} {104}},\ \bibinfo {pages} {090602} (\bibinfo
  {year} {2010})}\BibitemShut {NoStop}%
\bibitem [{\citenamefont {Bochkov}\ and\ \citenamefont
  {Kuzovlev}(1977{\natexlab{a}})}]{Bochkov_1977}%
  \BibitemOpen
  \bibfield  {author} {\bibinfo {author} {\bibfnamefont {G.~N.}\ \bibnamefont
  {Bochkov}}\ and\ \bibinfo {author} {\bibfnamefont {Y.~E.}\ \bibnamefont
  {Kuzovlev}},\ }\href@noop {} {\bibfield  {journal} {\bibinfo  {journal} {Zh.
  Eksp. Teor. Fiz.}\ }\textbf {\bibinfo {volume} {72}},\ \bibinfo {pages} {238}
  (\bibinfo {year} {1977}{\natexlab{a}})}\BibitemShut {NoStop}%
\bibitem [{\citenamefont {Bochkov}\ and\ \citenamefont
  {Kuzovlev}(1977{\natexlab{b}})}]{Bochkov_2_1977}%
  \BibitemOpen
  \bibfield  {author} {\bibinfo {author} {\bibfnamefont {G.~N.}\ \bibnamefont
  {Bochkov}}\ and\ \bibinfo {author} {\bibfnamefont {Y.~E.}\ \bibnamefont
  {Kuzovlev}},\ }\href@noop {} {\bibfield  {journal} {\bibinfo  {journal} {Sov.
  Phys.—JETP}\ }\textbf {\bibinfo {volume} {45}},\ \bibinfo {pages} {125}
  (\bibinfo {year} {1977}{\natexlab{b}})}\BibitemShut {NoStop}%
\bibitem [{\citenamefont {Bochkov}\ and\ \citenamefont
  {Kuzovlev}(1981{\natexlab{a}})}]{Bochkov_1981}%
  \BibitemOpen
  \bibfield  {author} {\bibinfo {author} {\bibfnamefont {G.~N.}\ \bibnamefont
  {Bochkov}}\ and\ \bibinfo {author} {\bibfnamefont {Y.~E.}\ \bibnamefont
  {Kuzovlev}},\ }\href@noop {} {\bibfield  {journal} {\bibinfo  {journal}
  {Physica}\ }\textbf {\bibinfo {volume} {106}},\ \bibinfo {pages} {443}
  (\bibinfo {year} {1981}{\natexlab{a}})}\BibitemShut {NoStop}%
\bibitem [{\citenamefont {Bochkov}\ and\ \citenamefont
  {Kuzovlev}(1981{\natexlab{b}})}]{Bochkov_2_1981}%
  \BibitemOpen
  \bibfield  {author} {\bibinfo {author} {\bibfnamefont {G.~N.}\ \bibnamefont
  {Bochkov}}\ and\ \bibinfo {author} {\bibfnamefont {Y.~E.}\ \bibnamefont
  {Kuzovlev}},\ }\href@noop {} {\bibfield  {journal} {\bibinfo  {journal}
  {Physica}\ }\textbf {\bibinfo {volume} {106}},\ \bibinfo {pages} {480}
  (\bibinfo {year} {1981}{\natexlab{b}})}\BibitemShut {NoStop}%
\bibitem [{\citenamefont {Bochkov}\ and\ \citenamefont
  {Kuzovlev}(2013)}]{Bochkov2013}%
  \BibitemOpen
  \bibfield  {author} {\bibinfo {author} {\bibfnamefont {G.~N.}\ \bibnamefont
  {Bochkov}}\ and\ \bibinfo {author} {\bibfnamefont {Y.~E.}\ \bibnamefont
  {Kuzovlev}},\ }\href@noop {} {\bibfield  {journal} {\bibinfo  {journal}
  {{Physics-Uspekhi}}\ }\textbf {\bibinfo {volume} {{56}}},\ \bibinfo {pages}
  {59} (\bibinfo {year} {{2013}})}\BibitemShut {NoStop}%
\bibitem [{\citenamefont {Jarzynski}(2007)}]{Jarzynski_comparison_2007}%
  \BibitemOpen
  \bibfield  {author} {\bibinfo {author} {\bibfnamefont {C.}~\bibnamefont
  {Jarzynski}},\ }\href@noop {} {\bibfield  {journal} {\bibinfo  {journal} {C.
  R. Physique}\ }\textbf {\bibinfo {volume} {8}},\ \bibinfo {pages} {495}
  (\bibinfo {year} {2007})}\BibitemShut {NoStop}%
\bibitem [{\citenamefont {Seifert}(2008)}]{Seifert_Stochastic_2008}%
  \BibitemOpen
  \bibfield  {author} {\bibinfo {author} {\bibfnamefont {U.}~\bibnamefont
  {Seifert}},\ }\href@noop {} {\bibfield  {journal} {\bibinfo  {journal} {Eur.
  Phys. J. B}\ }\textbf {\bibinfo {volume} {64}},\ \bibinfo {pages} {423}
  (\bibinfo {year} {2008})}\BibitemShut {NoStop}%
\bibitem [{\citenamefont {Campisi}\ \emph {et~al.}(2011)\citenamefont
  {Campisi}, \citenamefont {Hänggi},\ and\ \citenamefont
  {Talkner}}]{Campisi_2011}%
  \BibitemOpen
  \bibfield  {author} {\bibinfo {author} {\bibfnamefont {M.}~\bibnamefont
  {Campisi}}, \bibinfo {author} {\bibfnamefont {P.}~\bibnamefont {Hänggi}}, \
  and\ \bibinfo {author} {\bibfnamefont {P.}~\bibnamefont {Talkner}},\
  }\href@noop {} {\bibfield  {journal} {\bibinfo  {journal} {Rev. Mod. Phys.}\
  }\textbf {\bibinfo {volume} {83}},\ \bibinfo {pages} {771} (\bibinfo {year}
  {2011})}\BibitemShut {NoStop}%
\bibitem [{\citenamefont {Horowitz}\ and\ \citenamefont
  {Jarzynski}(2007)}]{Horowitz_2007}%
  \BibitemOpen
  \bibfield  {author} {\bibinfo {author} {\bibfnamefont {J.~M.}\ \bibnamefont
  {Horowitz}}\ and\ \bibinfo {author} {\bibfnamefont {C.}~\bibnamefont
  {Jarzynski}},\ }\href@noop {} {\bibfield  {journal} {\bibinfo  {journal} {J.
  Stat. Mech.}\ ,\ \bibinfo {pages} {P11002}} (\bibinfo {year}
  {2007})}\BibitemShut {NoStop}%
\bibitem [{\citenamefont {Sivak}\ and\ \citenamefont
  {Crooks}(2012{\natexlab{a}})}]{Sivak2012}%
  \BibitemOpen
  \bibfield  {author} {\bibinfo {author} {\bibfnamefont {D.~A.}\ \bibnamefont
  {Sivak}}\ and\ \bibinfo {author} {\bibfnamefont {G.~E.}\ \bibnamefont
  {Crooks}},\ }\href@noop {} {\bibfield  {journal} {\bibinfo  {journal} {{Phys.
  Rev. Lett.}}\ }\textbf {\bibinfo {volume} {{108}}},\ \bibinfo {pages}
  {190602} (\bibinfo {year} {{2012}}{\natexlab{a}})}\BibitemShut {NoStop}%
\bibitem [{\citenamefont {Zulkowski}\ \emph {et~al.}(2012)\citenamefont
  {Zulkowski}, \citenamefont {Sivak}, \citenamefont {Crooks},\ and\
  \citenamefont {DeWeese}}]{Zulkowski2012}%
  \BibitemOpen
  \bibfield  {author} {\bibinfo {author} {\bibfnamefont {P.~R.}\ \bibnamefont
  {Zulkowski}}, \bibinfo {author} {\bibfnamefont {D.~A.}\ \bibnamefont
  {Sivak}}, \bibinfo {author} {\bibfnamefont {G.~E.}\ \bibnamefont {Crooks}}, \
  and\ \bibinfo {author} {\bibfnamefont {M.~R.}\ \bibnamefont {DeWeese}},\
  }\href@noop {} {\bibfield  {journal} {\bibinfo  {journal} {{Phys. Rev. E}}\
  }\textbf {\bibinfo {volume} {{86}}},\ \bibinfo {pages} {041148} (\bibinfo
  {year} {{2012}})}\BibitemShut {NoStop}%
\bibitem [{\citenamefont {Zulkowski}\ \emph {et~al.}(2013)\citenamefont
  {Zulkowski}, \citenamefont {Sivak},\ and\ \citenamefont
  {DeWeese}}]{Zulkowski_PLOS_ONE_2013}%
  \BibitemOpen
  \bibfield  {author} {\bibinfo {author} {\bibfnamefont {P.~R.}\ \bibnamefont
  {Zulkowski}}, \bibinfo {author} {\bibfnamefont {D.~A.}\ \bibnamefont
  {Sivak}}, \ and\ \bibinfo {author} {\bibfnamefont {M.~R.}\ \bibnamefont
  {DeWeese}},\ }\href@noop {} {\bibfield  {journal} {\bibinfo  {journal}
  {Public Library of Science}\ }\textbf {\bibinfo {volume} {(in press)}}
  (\bibinfo {year} {2013})}\BibitemShut {NoStop}%
\bibitem [{\citenamefont {Shenfeld}\ \emph {et~al.}(2009)\citenamefont
  {Shenfeld}, \citenamefont {Xu}, \citenamefont {Eastwood}, \citenamefont
  {Dror},\ and\ \citenamefont {Shaw}}]{Shenfeld2009}%
  \BibitemOpen
  \bibfield  {author} {\bibinfo {author} {\bibfnamefont {D.~K.}\ \bibnamefont
  {Shenfeld}}, \bibinfo {author} {\bibfnamefont {H.}~\bibnamefont {Xu}},
  \bibinfo {author} {\bibfnamefont {M.~P.}\ \bibnamefont {Eastwood}}, \bibinfo
  {author} {\bibfnamefont {R.~O.}\ \bibnamefont {Dror}}, \ and\ \bibinfo
  {author} {\bibfnamefont {D.~E.}\ \bibnamefont {Shaw}},\ }\href@noop {}
  {\bibfield  {journal} {\bibinfo  {journal} {Phys. Rev. E}\ }\textbf {\bibinfo
  {volume} {80}},\ \bibinfo {pages} {046705} (\bibinfo {year}
  {2009})}\BibitemShut {NoStop}%
\bibitem [{\citenamefont {Brody}\ and\ \citenamefont {Hook}(2009)}]{Brody2009}%
  \BibitemOpen
  \bibfield  {author} {\bibinfo {author} {\bibfnamefont {D.~C.}\ \bibnamefont
  {Brody}}\ and\ \bibinfo {author} {\bibfnamefont {D.~W.}\ \bibnamefont
  {Hook}},\ }\href@noop {} {\bibfield  {journal} {\bibinfo  {journal} {J. Phys.
  A}\ }\textbf {\bibinfo {volume} {42}},\ \bibinfo {pages} {023001} (\bibinfo
  {year} {2009})}\BibitemShut {NoStop}%
\bibitem [{\citenamefont {Gomez-Marin}\ \emph {et~al.}(2008)\citenamefont
  {Gomez-Marin}, \citenamefont {Schmiedl},\ and\ \citenamefont
  {Seifert}}]{Seifert2008}%
  \BibitemOpen
  \bibfield  {author} {\bibinfo {author} {\bibfnamefont {A.}~\bibnamefont
  {Gomez-Marin}}, \bibinfo {author} {\bibfnamefont {T.}~\bibnamefont
  {Schmiedl}}, \ and\ \bibinfo {author} {\bibfnamefont {U.}~\bibnamefont
  {Seifert}},\ }\href@noop {} {\bibfield  {journal} {\bibinfo  {journal} {J.
  Chem. Phys.}\ }\textbf {\bibinfo {volume} {129}},\ \bibinfo {pages} {024114
  (8)} (\bibinfo {year} {2008})}\BibitemShut {NoStop}%
\bibitem [{\citenamefont {Schmiedl}\ and\ \citenamefont
  {Seifert}(2007)}]{Seifert2007}%
  \BibitemOpen
  \bibfield  {author} {\bibinfo {author} {\bibfnamefont {T.}~\bibnamefont
  {Schmiedl}}\ and\ \bibinfo {author} {\bibfnamefont {U.}~\bibnamefont
  {Seifert}},\ }\href@noop {} {\bibfield  {journal} {\bibinfo  {journal} {Phys.
  Rev. Lett.}\ }\textbf {\bibinfo {volume} {98}},\ \bibinfo {pages} {108301}
  (\bibinfo {year} {2007})}\BibitemShut {NoStop}%
\bibitem [{\citenamefont {Aurell}\ \emph {et~al.}(2011)\citenamefont {Aurell},
  \citenamefont {Mej\'{i}a-Monasterio},\ and\ \citenamefont
  {Muratore-Ginanneschi}}]{Aurell2011}%
  \BibitemOpen
  \bibfield  {author} {\bibinfo {author} {\bibfnamefont {E.}~\bibnamefont
  {Aurell}}, \bibinfo {author} {\bibfnamefont {C.}~\bibnamefont
  {Mej\'{i}a-Monasterio}}, \ and\ \bibinfo {author} {\bibfnamefont
  {P.}~\bibnamefont {Muratore-Ginanneschi}},\ }\href@noop {} {\bibfield
  {journal} {\bibinfo  {journal} {Phys. Rev. Lett.}\ }\textbf {\bibinfo
  {volume} {106}},\ \bibinfo {pages} {250601 (4)} (\bibinfo {year}
  {2011})}\BibitemShut {NoStop}%
\bibitem [{\citenamefont {Andresen}(2011)}]{Andresen2011}%
  \BibitemOpen
  \bibfield  {author} {\bibinfo {author} {\bibfnamefont {B.}~\bibnamefont
  {Andresen}},\ }\href@noop {} {\bibfield  {journal} {\bibinfo  {journal}
  {Angew. Chem., Int. Ed.}\ }\textbf {\bibinfo {volume} {50}},\ \bibinfo
  {pages} {2690} (\bibinfo {year} {2011})}\BibitemShut {NoStop}%
\bibitem [{\citenamefont {Chen}\ and\ \citenamefont {Sun}(2004)}]{Chen2004}%
  \BibitemOpen
  \bibinfo {editor} {\bibfnamefont {L.}~\bibnamefont {Chen}}\ and\ \bibinfo
  {editor} {\bibfnamefont {F.}~\bibnamefont {Sun}},\ eds.,\ \href@noop {}
  {\emph {\bibinfo {title} {Advances in finite time thermodynamics: analysis
  and optimization}}}\ (\bibinfo  {publisher} {Nova Science Publishers},\
  \bibinfo {address} {New York},\ \bibinfo {year} {2004})\BibitemShut {NoStop}%
\bibitem [{\citenamefont {Alberts}\ \emph {et~al.}(2002)\citenamefont
  {Alberts}, \citenamefont {Johnson}, \citenamefont {Lewis}, \citenamefont
  {Raff}, \citenamefont {Roberts},\ and\ \citenamefont {Walter}}]{MBOC}%
  \BibitemOpen
  \bibfield  {author} {\bibinfo {author} {\bibfnamefont {B.}~\bibnamefont
  {Alberts}}, \bibinfo {author} {\bibfnamefont {A.}~\bibnamefont {Johnson}},
  \bibinfo {author} {\bibfnamefont {J.}~\bibnamefont {Lewis}}, \bibinfo
  {author} {\bibfnamefont {M.}~\bibnamefont {Raff}}, \bibinfo {author}
  {\bibfnamefont {K.}~\bibnamefont {Roberts}}, \ and\ \bibinfo {author}
  {\bibfnamefont {P.}~\bibnamefont {Walter}},\ }\href@noop {} {\emph {\bibinfo
  {title} {Molecular Biology of the Cell}}}\ (\bibinfo  {publisher} {Garland
  Science},\ \bibinfo {address} {New York},\ \bibinfo {year}
  {2002})\BibitemShut {NoStop}%
\bibitem [{\citenamefont {Zulkowski}\ and\ \citenamefont
  {DeWeese}(2014)}]{Zulkowski2014}%
  \BibitemOpen
  \bibfield  {author} {\bibinfo {author} {\bibfnamefont {P.~R.}\ \bibnamefont
  {Zulkowski}}\ and\ \bibinfo {author} {\bibfnamefont {M.~R.}\ \bibnamefont
  {DeWeese}},\ }\href@noop {} {\bibfield  {journal} {\bibinfo  {journal}
  {{Phys. Rev. E}}\ }\textbf {\bibinfo {volume} {{89}}},\ \bibinfo {pages}
  {052140} (\bibinfo {year} {{2014}})}\BibitemShut {NoStop}%
\bibitem [{\citenamefont {Weinhold}(1975)}]{Weinhold1975a}%
  \BibitemOpen
  \bibfield  {author} {\bibinfo {author} {\bibfnamefont {F.}~\bibnamefont
  {Weinhold}},\ }\href@noop {} {\bibfield  {journal} {\bibinfo  {journal} {J.
  Chem. Phys.}\ }\textbf {\bibinfo {volume} {63}},\ \bibinfo {pages} {2479}
  (\bibinfo {year} {1975})}\BibitemShut {NoStop}%
\bibitem [{\citenamefont {Ruppeiner}(1979)}]{Ruppeiner1979}%
  \BibitemOpen
  \bibfield  {author} {\bibinfo {author} {\bibfnamefont {G.}~\bibnamefont
  {Ruppeiner}},\ }\href@noop {} {\bibfield  {journal} {\bibinfo  {journal}
  {Phys. Rev. A}\ }\textbf {\bibinfo {volume} {20}},\ \bibinfo {pages} {1608}
  (\bibinfo {year} {1979})}\BibitemShut {NoStop}%
\bibitem [{\citenamefont {Schl\"ogl}(1985)}]{Schlogl1985}%
  \BibitemOpen
  \bibfield  {author} {\bibinfo {author} {\bibfnamefont {F.}~\bibnamefont
  {Schl\"ogl}},\ }\href@noop {} {\bibfield  {journal} {\bibinfo  {journal} {Z.
  Phys. B}\ }\textbf {\bibinfo {volume} {59}},\ \bibinfo {pages} {449}
  (\bibinfo {year} {1985})}\BibitemShut {NoStop}%
\bibitem [{\citenamefont {Salamon}\ \emph {et~al.}(1984)\citenamefont
  {Salamon}, \citenamefont {Nulton},\ and\ \citenamefont
  {Ihrig}}]{Salamon1984}%
  \BibitemOpen
  \bibfield  {author} {\bibinfo {author} {\bibfnamefont {P.}~\bibnamefont
  {Salamon}}, \bibinfo {author} {\bibfnamefont {J.}~\bibnamefont {Nulton}}, \
  and\ \bibinfo {author} {\bibfnamefont {E.}~\bibnamefont {Ihrig}},\
  }\href@noop {} {\bibfield  {journal} {\bibinfo  {journal} {J. Chem. Phys.}\
  }\textbf {\bibinfo {volume} {80}},\ \bibinfo {pages} {436} (\bibinfo {year}
  {1984})}\BibitemShut {NoStop}%
\bibitem [{\citenamefont {Salamon}\ and\ \citenamefont
  {Berry}(1983)}]{Salamon1983a}%
  \BibitemOpen
  \bibfield  {author} {\bibinfo {author} {\bibfnamefont {P.}~\bibnamefont
  {Salamon}}\ and\ \bibinfo {author} {\bibfnamefont {R.~S.}\ \bibnamefont
  {Berry}},\ }\href@noop {} {\bibfield  {journal} {\bibinfo  {journal} {Phys.
  Rev. Lett.}\ }\textbf {\bibinfo {volume} {51}},\ \bibinfo {pages} {1127}
  (\bibinfo {year} {1983})}\BibitemShut {NoStop}%
\bibitem [{\citenamefont {Brody}\ and\ \citenamefont
  {Rivier}(1995)}]{Brody1995}%
  \BibitemOpen
  \bibfield  {author} {\bibinfo {author} {\bibfnamefont {D.}~\bibnamefont
  {Brody}}\ and\ \bibinfo {author} {\bibfnamefont {N.}~\bibnamefont {Rivier}},\
  }\href@noop {} {\bibfield  {journal} {\bibinfo  {journal} {Phys. Rev. E}\
  }\textbf {\bibinfo {volume} {51}},\ \bibinfo {pages} {1006} (\bibinfo {year}
  {1995})}\BibitemShut {NoStop}%
\bibitem [{\citenamefont {Crooks}(2007)}]{Crooks2007c}%
  \BibitemOpen
  \bibfield  {author} {\bibinfo {author} {\bibfnamefont {G.~E.}\ \bibnamefont
  {Crooks}},\ }\href@noop {} {\bibfield  {journal} {\bibinfo  {journal} {Phys.
  Rev. Lett.}\ }\textbf {\bibinfo {volume} {99}},\ \bibinfo {pages} {100602}
  (\bibinfo {year} {2007})}\BibitemShut {NoStop}%
\bibitem [{\citenamefont {Burbea}\ and\ \citenamefont
  {Rao}(1982)}]{Burbea1982}%
  \BibitemOpen
  \bibfield  {author} {\bibinfo {author} {\bibfnamefont {J.}~\bibnamefont
  {Burbea}}\ and\ \bibinfo {author} {\bibfnamefont {C.~R.}\ \bibnamefont
  {Rao}},\ }\href@noop {} {\bibfield  {journal} {\bibinfo  {journal} {J.
  Multivariate Anal.}\ }\textbf {\bibinfo {volume} {12}},\ \bibinfo {pages}
  {575} (\bibinfo {year} {1982})}\BibitemShut {NoStop}%
\bibitem [{\citenamefont {Mandal}\ and\ \citenamefont
  {Jarzynski}(2015)}]{Mandal2015}%
  \BibitemOpen
  \bibfield  {author} {\bibinfo {author} {\bibfnamefont {D.}~\bibnamefont
  {Mandal}}\ and\ \bibinfo {author} {\bibfnamefont {C.}~\bibnamefont
  {Jarzynski}},\ }\href@noop {} {\enquote {\bibinfo {title} {Analysis of slow
  transitions between nonequilibrium steady states},}\ } (\bibinfo {year}
  {2015}),\ \bibinfo {note} {arXiv:1507.06269}\BibitemShut {NoStop}%
\bibitem [{\citenamefont {Zwanzig}(2001)}]{Zwanzig2001}%
  \BibitemOpen
  \bibfield  {author} {\bibinfo {author} {\bibfnamefont {R.}~\bibnamefont
  {Zwanzig}},\ }\href@noop {} {\emph {\bibinfo {title} {Nonequilibrium
  statistical mechanics}}}\ (\bibinfo  {publisher} {Oxford University Press},\
  \bibinfo {address} {New York},\ \bibinfo {year} {2001})\BibitemShut {NoStop}%
\bibitem [{\citenamefont {Dillenschneider}\ and\ \citenamefont
  {Lutz}(2009)}]{Lutz2009}%
  \BibitemOpen
  \bibfield  {author} {\bibinfo {author} {\bibfnamefont {R.}~\bibnamefont
  {Dillenschneider}}\ and\ \bibinfo {author} {\bibfnamefont {E.}~\bibnamefont
  {Lutz}},\ }\href@noop {} {\bibfield  {journal} {\bibinfo  {journal} {{Phys.
  Rev. Lett.}}\ }\textbf {\bibinfo {volume} {{102}}},\ \bibinfo {pages}
  {210601} (\bibinfo {year} {{2009}})}\BibitemShut {NoStop}%
\bibitem [{\citenamefont {Beacuterut}\ \emph {et~al.}(2012)\citenamefont
  {Beacuterut}, \citenamefont {Arakelyan}, \citenamefont {Petrosyan},
  \citenamefont {Ciliberto}, \citenamefont {Dillenschneider},\ and\
  \citenamefont {Lutz}}]{Lutz2012}%
  \BibitemOpen
  \bibfield  {author} {\bibinfo {author} {\bibfnamefont {A.}~\bibnamefont
  {Beacuterut}}, \bibinfo {author} {\bibfnamefont {A.}~\bibnamefont
  {Arakelyan}}, \bibinfo {author} {\bibfnamefont {A.}~\bibnamefont
  {Petrosyan}}, \bibinfo {author} {\bibfnamefont {S.}~\bibnamefont
  {Ciliberto}}, \bibinfo {author} {\bibfnamefont {R.}~\bibnamefont
  {Dillenschneider}}, \ and\ \bibinfo {author} {\bibfnamefont {E.}~\bibnamefont
  {Lutz}},\ }\href@noop {} {\bibfield  {journal} {\bibinfo  {journal}
  {{Nature}}\ }\textbf {\bibinfo {volume} {{483}}},\ \bibinfo {pages} {187}
  (\bibinfo {year} {{2012}})}\BibitemShut {NoStop}%
\bibitem [{\citenamefont {Reimann}(2002)}]{Reimann2002}%
  \BibitemOpen
  \bibfield  {author} {\bibinfo {author} {\bibfnamefont {P.}~\bibnamefont
  {Reimann}},\ }\href {\doibase
  http://dx.doi.org/10.1016/S0370-1573(01)00081-3} {\bibfield  {journal}
  {\bibinfo  {journal} {Physics Reports}\ }\textbf {\bibinfo {volume} {361}},\
  \bibinfo {pages} {57 } (\bibinfo {year} {2002})}\BibitemShut {NoStop}%
\bibitem [{\citenamefont {Sivak}\ and\ \citenamefont
  {Crooks}(2012{\natexlab{b}})}]{Sivak2012b}%
  \BibitemOpen
  \bibfield  {author} {\bibinfo {author} {\bibfnamefont {D.~A.}\ \bibnamefont
  {Sivak}}\ and\ \bibinfo {author} {\bibfnamefont {G.~E.}\ \bibnamefont
  {Crooks}},\ }\href@noop {} {\bibfield  {journal} {\bibinfo  {journal} {Phys.
  Rev. Lett.}\ }\textbf {\bibinfo {volume} {108}},\ \bibinfo {pages} {190602}
  (\bibinfo {year} {2012}{\natexlab{b}})}\BibitemShut {NoStop}%
\bibitem [{\citenamefont {Pankratov}(1999)}]{Pankratov1999}%
  \BibitemOpen
  \bibfield  {author} {\bibinfo {author} {\bibfnamefont {A.}~\bibnamefont
  {Pankratov}},\ }\href@noop {} {\bibfield  {journal} {\bibinfo  {journal}
  {Phys. Lett. A}\ }\textbf {\bibinfo {volume} {255}},\ \bibinfo {pages} {17}
  (\bibinfo {year} {1999})}\BibitemShut {NoStop}%
\bibitem [{\citenamefont {Berezhkovskii}\ and\ \citenamefont
  {Szabo}(2011)}]{Szabo2011}%
  \BibitemOpen
  \bibfield  {author} {\bibinfo {author} {\bibfnamefont {A.}~\bibnamefont
  {Berezhkovskii}}\ and\ \bibinfo {author} {\bibfnamefont {A.}~\bibnamefont
  {Szabo}},\ }\href@noop {} {\bibfield  {journal} {\bibinfo  {journal}
  {{Journal of Chemical Physics}}\ }\textbf {\bibinfo {volume} {{135}}},\
  \bibinfo {pages} {074108 (5 pp.)} (\bibinfo {year} {{2011}})}\BibitemShut
  {NoStop}%
\bibitem [{\citenamefont {Seifert}(2012)}]{SeifertReview_2012}%
  \BibitemOpen
  \bibfield  {author} {\bibinfo {author} {\bibfnamefont {U.}~\bibnamefont
  {Seifert}},\ }\href {http://stacks.iop.org/0034-4885/75/i=12/a=126001}
  {\bibfield  {journal} {\bibinfo  {journal} {Reports on Progress in Physics}\
  }\textbf {\bibinfo {volume} {75}},\ \bibinfo {pages} {126001} (\bibinfo
  {year} {2012})}\BibitemShut {NoStop}%
\bibitem [{\citenamefont {Then}\ and\ \citenamefont {Engel}(2008)}]{Then2008}%
  \BibitemOpen
  \bibfield  {author} {\bibinfo {author} {\bibfnamefont {H.}~\bibnamefont
  {Then}}\ and\ \bibinfo {author} {\bibfnamefont {A.}~\bibnamefont {Engel}},\
  }\href@noop {} {\bibfield  {journal} {\bibinfo  {journal} {{Phys. Rev. E}}\
  }\textbf {\bibinfo {volume} {{77}}},\ \bibinfo {pages} {{041105}} (\bibinfo
  {year} {{2008}})}\BibitemShut {NoStop}%
\bibitem [{\citenamefont {Esposito}\ \emph {et~al.}(2010)\citenamefont
  {Esposito}, \citenamefont {Kawai}, \citenamefont {Lindenberg},\ and\
  \citenamefont {Van~den Broeck}}]{Esposito2010}%
  \BibitemOpen
  \bibfield  {author} {\bibinfo {author} {\bibfnamefont {M.}~\bibnamefont
  {Esposito}}, \bibinfo {author} {\bibfnamefont {R.}~\bibnamefont {Kawai}},
  \bibinfo {author} {\bibfnamefont {K.}~\bibnamefont {Lindenberg}}, \ and\
  \bibinfo {author} {\bibfnamefont {C.}~\bibnamefont {Van~den Broeck}},\
  }\href@noop {} {\bibfield  {journal} {\bibinfo  {journal} {{Europhys.
  Lett.}}\ }\textbf {\bibinfo {volume} {{89}}},\ \bibinfo {pages} {20003}
  (\bibinfo {year} {{2010}})}\BibitemShut {NoStop}%
\bibitem [{\citenamefont {Aurell}\ \emph {et~al.}(2012)\citenamefont {Aurell},
  \citenamefont {Gawedzki}, \citenamefont {Mejia-Monasterio}, \citenamefont
  {Mohayaee},\ and\ \citenamefont {Muratore-Ginanneschi}}]{Aurell2012}%
  \BibitemOpen
  \bibfield  {author} {\bibinfo {author} {\bibfnamefont {E.}~\bibnamefont
  {Aurell}}, \bibinfo {author} {\bibfnamefont {K.}~\bibnamefont {Gawedzki}},
  \bibinfo {author} {\bibfnamefont {C.}~\bibnamefont {Mejia-Monasterio}},
  \bibinfo {author} {\bibfnamefont {R.}~\bibnamefont {Mohayaee}}, \ and\
  \bibinfo {author} {\bibfnamefont {P.}~\bibnamefont {Muratore-Ginanneschi}},\
  }\href@noop {} {\bibfield  {journal} {\bibinfo  {journal} {{J. Stat. Phys.}}\
  }\textbf {\bibinfo {volume} {{147}}},\ \bibinfo {pages} {487} (\bibinfo
  {year} {{2012}})}\BibitemShut {NoStop}%
\bibitem [{\citenamefont {Diana}\ \emph {et~al.}(2013)\citenamefont {Diana},
  \citenamefont {Bagci},\ and\ \citenamefont {Esposito}}]{Diana2013}%
  \BibitemOpen
  \bibfield  {author} {\bibinfo {author} {\bibfnamefont {G.}~\bibnamefont
  {Diana}}, \bibinfo {author} {\bibfnamefont {G.~B.}\ \bibnamefont {Bagci}}, \
  and\ \bibinfo {author} {\bibfnamefont {M.}~\bibnamefont {Esposito}},\
  }\href@noop {} {\bibfield  {journal} {\bibinfo  {journal} {{Phys. Rev. E}}\
  }\textbf {\bibinfo {volume} {{87}}},\ \bibinfo {pages} {012111} (\bibinfo
  {year} {{2013}})}\BibitemShut {NoStop}%
\bibitem [{\citenamefont {Plenio}\ and\ \citenamefont
  {Vitelli}(2001)}]{Plenio2001}%
  \BibitemOpen
  \bibfield  {author} {\bibinfo {author} {\bibfnamefont {M.}~\bibnamefont
  {Plenio}}\ and\ \bibinfo {author} {\bibfnamefont {V.}~\bibnamefont
  {Vitelli}},\ }\href@noop {} {\bibfield  {journal} {\bibinfo  {journal}
  {{Contemp. Phys.}}\ }\textbf {\bibinfo {volume} {{42}}},\ \bibinfo {pages}
  {25} (\bibinfo {year} {{2001}})}\BibitemShut {NoStop}%
\bibitem [{\citenamefont {Abramowitz}\ and\ \citenamefont
  {Stegun}(1972)}]{Stegun1972}%
  \BibitemOpen
  \bibfield  {author} {\bibinfo {author} {\bibfnamefont {M.}~\bibnamefont
  {Abramowitz}}\ and\ \bibinfo {author} {\bibfnamefont {I.}~\bibnamefont
  {Stegun}},\ }\href@noop {} {\emph {\bibinfo {title} {Handbook of Mathematical
  Functions with Formulas, Graphs, and Mathematical Tables}}}\ (\bibinfo
  {publisher} {Dover Publications, New York},\ \bibinfo {year}
  {1972})\BibitemShut {NoStop}%
\bibitem [{\citenamefont {Likharev}(1982)}]{Likharev1982}%
  \BibitemOpen
  \bibfield  {author} {\bibinfo {author} {\bibfnamefont {K.}~\bibnamefont
  {Likharev}},\ }in\ \href@noop {} {\emph {\bibinfo {booktitle} {{International
  Journal of Theoretical Physics}}}},\ Vol.~\bibinfo {volume} {{21}}\ (\bibinfo
  {organization} {{MIT Lab. Comput. Sci.; Army Res. Office; IBM Corp.; NSF;
  XEROX Corp.}},\ \bibinfo {address} {{UK}},\ \bibinfo {year} {{1982}})\ pp.\
  \bibinfo {pages} {{311--26}},\ \bibinfo {note} {{`Physics of Computation'
  Conference, 6-8 May 1981, Dedham, MA, USA}}\BibitemShut {NoStop}%
\bibitem [{\citenamefont {Pankratov}(1997)}]{Pankratov1997}%
  \BibitemOpen
  \bibfield  {author} {\bibinfo {author} {\bibfnamefont {A.}~\bibnamefont
  {Pankratov}},\ }\href@noop {} {\bibfield  {journal} {\bibinfo  {journal}
  {{Physics Letters A}}\ }\textbf {\bibinfo {volume} {{234}}},\ \bibinfo
  {pages} {329} (\bibinfo {year} {{1997}})}\BibitemShut {NoStop}%
\bibitem [{\citenamefont {{Nikitenkova, Svetlana P. and Pankratov, Andrey
  L.}}(1998)}]{Nikitenkova1998}%
  \BibitemOpen
  \bibfield  {author} {\bibinfo {author} {\bibnamefont {{Nikitenkova, Svetlana
  P. and Pankratov, Andrey L.}}},\ }\href {\doibase {10.1103/PhysRevE.58.6964}}
  {\bibfield  {journal} {\bibinfo  {journal} {{Phys. Rev. E}}\ }\textbf
  {\bibinfo {volume} {{58}}},\ \bibinfo {pages} {6964} (\bibinfo {year}
  {{1998}})}\BibitemShut {NoStop}%
\bibitem [{\citenamefont {{Yonggun Jun}}\ \emph {et~al.}(2014)\citenamefont
  {{Yonggun Jun}}, \citenamefont {Gavrilov},\ and\ \citenamefont
  {Bechhoefer}}]{Jun2014}%
  \BibitemOpen
  \bibfield  {author} {\bibinfo {author} {\bibnamefont {{Yonggun Jun}}},
  \bibinfo {author} {\bibfnamefont {M.}~\bibnamefont {Gavrilov}}, \ and\
  \bibinfo {author} {\bibfnamefont {J.}~\bibnamefont {Bechhoefer}},\
  }\href@noop {} {\bibfield  {journal} {\bibinfo  {journal} {{Physical Review
  Letters}}\ }\textbf {\bibinfo {volume} {{113}}},\ \bibinfo {pages} {190601 (5
  pp.)} (\bibinfo {year} {{2014}})}\BibitemShut {NoStop}%
\end{thebibliography}%

\end{document}